\documentclass[prd,twocolumn,showpacs,floatfix,amsmath,nofootinbib,amssymb,floatfix]{revtex4}
\usepackage{graphicx,color,dcolumn,booktabs,bm}
\usepackage{longtable,lscape}
\usepackage{txfonts}
\usepackage{overpic}
\usepackage{amssymb}
\usepackage{indentfirst}
\usepackage{feynmf}   %{feynmp}
\usepackage{slashed}  %for Feynman symbols
\usepackage{cases}
\usepackage{color}
\usepackage{multirow}
\usepackage{epstopdf}
\usepackage{graphicx,color,dcolumn,booktabs,bm}
\usepackage[colorlinks,
            citecolor=blue,
            anchorcolor=red,
            menucolor=red,
            linkcolor=red,
            filecolor=red,
            runcolor=red,
            urlcolor=blue,
            frenchlinks=red]{hyperref}

\graphicspath{{Figures/}} %

\allowdisplaybreaks

\begin{document}
%\begin{CJK}{GBK}{}

\title{Predictions of the hidden-charm molecular states with the four-quark component}

\author{Rui Chen$^{1,2}$}
\email{chenr2012@lzu.edu.cn}
\author{Xiang Liu$^{1,2}$}
\email{xiangliu@lzu.edu.cn}
\author{Yan-Rui Liu$^{3,4}$}
\email{yrliu@sdu.edu.cn}
\author{Shi-Lin Zhu$^{5,6,7}$}
\email{zhusl@pku.edu.cn} \affiliation{ $^1$School of Physical
Science and Technology, Lanzhou University, Lanzhou 730000, China
\\
$^2$Research Center for Hadron and CSR Physics, Lanzhou University
and Institute of Modern Physics of CAS, Lanzhou 730000, China
\\
$^3$School of Physics and Key
Laboratory of Particle Physics and Particle Irradiation (MOE), Shandong University, Jinan 250100, China\\
$^4$Key Laboratory of Theoretical Physics, Institute of
Theoretical Physics, CAS, Beijing 100190, China
\\
$^5$School of Physics and State Key Laboratory of Nuclear Physics
and Technology, Peking University, Beijing 100871, China
\\
$^6$Collaborative Innovation Center of Quantum Matter, Beijing
100871, China
\\
$^7$Center of High Energy Physics, Peking University, Beijing
100871, China }

\begin{abstract}
In this work, we study the $T\bar{T}$-type molecular systems
systematically via one pion exchange model, where $T$ denotes
the narrow $J^P=1^+$ $D_1$ meson or $2^+$ $D_2^*$ meson and
$\bar{T}$ is its antiparticle. With the effective
potentials, we try to find the bound state solutions of the
corresponding systems, which provide crucial information of whether
there exist the $T\bar{T}$-type molecular states. According to our
analysis, we predict some $T\bar{T}$-type molecular states which may
be accessible at future experiments like LHCb and forthcoming
BelleII.
\end{abstract}


\maketitle

\section{Introduction}\label{sec1}

Searching for exotic states is a promising research topic full of
opportunities and challenges in hadron physics. In 2013, the
observation of the charged charmonium-like structure $Z_c(3900)$
from the BESIII \cite{Ablikim:2013mio} and Belle \cite{Liu:2013dau}
collaborations have inspired extensive discussions on the four-quark
matter. Very recently, the LHCb Collaboration announced two
hidden-charm pentaquarks $P_c(4380)$ and $P_c(4450)$
\cite{Aaij:2015tga}, which have aroused theorists' interest in the
five-quark matter again
\cite{Chen:2015loa,Chen:2015moa,Roca:2015dva,Mironov:2015ica,He:2015cea}.

As one of the possible exotic hadron configurations, the multiquark
state is a new hadronic matter beyond the conventional $q\bar{q}$
meson and $qqq$ baryon. Among the multiquark states, the molecular
state is a very popular configuration, which has been applied to
understand some of the experimental observations of the
near-threshold charmonium-like states, which are also named as the
$XYZ$ states.

In the past decade, the hidden-charm molecular states were studied
extensively, since these investigations have a close relation to
$X(3872)$
\cite{Liu:2008fh,Swanson:2003tb,Tornqvist:2004qy,Tornqvist:1993vu,Tornqvist:1993ng,Li:2012cs,Liu:2008tn},
$Y(3930)$ \cite{Liu:2008tn,Liu:2009uz,Branz:2009yt}, $Y(4140)$
\cite{Liu:2009ei,Liu:2008tn,Liu:2009uz,Branz:2009yt}, $Y(4274)$
\cite{Liu:2010hf,He:2011ed}, $Z_b(10610)$, $Z_b(10650)$
\cite{Liu:2008fh,Liu:2008tn,Sun:2011uh,Sun:2012zzd}, $Z^+(4430)$
\cite{Liu:2007bf,Liu:2008xz}, and newly observed $P_c(4380)$ and
$P_c(4450)$
\cite{Chen:2015loa,Chen:2015moa,Roca:2015dva,Mironov:2015ica,He:2015cea}.
As indicated in Refs. \cite{Li:2014gra,Karliner:2015ina}, the
hidden-charm quantum number may be a crucial condition for the
existence of the exotic molecular states. Thus, we can understand
naturally why so many charmonium-like states (or bottomonium-like)
states can be assigned into the hidden-charm molecular states.

With further experimental progress, more candidates of exotic states
will be reported. For theorists, it is time to make reliable
predictions of the hidden-charm molecular states. In this work, we
focus on the hidden-charm molecular states with four-quark
component, which are composed of the charmed and anti-charmed
mesons. Their properties are determined by the corresponding
components (charmed and anti-charmed mesons).

In the heavy quark limit \cite{Manohar:2000dt}, the S-wave and
P-wave charmed mesons can be grouped into three doublets
$H=(0^-,1^-)$, $S=(0^+,1^+)$, and $T=(1^+,2^+)$. In Ref.
\cite{Liu:2008tn}, the hidden-charm molecular states composed of the
charmed/anti-charmed mesons in the $H$ doublet were studied, which
are abbreviated as the $H\bar{H}$-type hidden-charm molecular
states. Later, the possible $S\bar{S}$-type hidden-charm molecular
states are predicted \cite{Hu:2010fg}, which are constructed by
$P$-wave charm-strange meson $D_{s0}(2317)$ and $D_{s1}(2460)$ and
their corresponding anti-particles, where $D_{s0}(2317)$ and
$D_{s1}(2460)$ have very narrow widths. In addition, the authors of
Ref. \cite{Shen:2010ky} analyzed the molecular systems composed of
the charmed mesons in the $H$ doublet and $S$ doublet carefully,
which are also named as $H\bar{S}$-type hidden-charm molecular
states.

Along this line, in this work, we continue to carry out the study of
the $T\bar{T}$-type hidden-charm molecular states, which are
composed of the charmed and anti-charmed mesons in the $T$ doublet.
The charmed mesons in the $T$ doublet have narrow widths. In Ref. \cite{Filin:2010se}, it is argued that the broad width of an open-charm meson probably results in unobservable bound states containing this meson. Thus, the
charmed mesons in the $T$ doublet are the suitable building blocks
of the molecular states.

In order to investigate the $T\bar{T}$-type hidden-charm molecular
states, we adopt the one pion exchange (OPE) model
\cite{Tornqvist:1993vu,Tornqvist:1993ng,Close:2009ag} to deduce the
effective potentials. With the obtained potentials and by solving
Schr\"odinger equation, we finally get the bound state solution and
can judge whether there exists the bound states. We need to specify
that the S-D mixing effect
\cite{Sun:2012zzd,Sun:2011uh,Yang:2011wz,Sun:2012sy} and
coupled-channel effect \cite{Li:2012ss,Chen:2014mwa} will be
included in our calculation. The detailed deductions of the OPE
potentials will be given in the next section. We hope that the
present study may stimulate interest in searching for the
$T\bar{T}$-type hidden-charm molecular states in experiments like
LHCb and forthcoming BelleII.

A hadron-hadron bound state such as the deuteron is stable since its
constituents do not decay. In the present case, both $D_1$ and
$D_2^*$ have a width of tens of MeV. Although a binding energy of
several MeV for a possible molecule is smaller than the width of its
constituent, some molecular type resonance may still result from the
$T\bar{T}$ interaction. This character may be approximately studied
by treating the mesons as stable states and encoding the width
effects into some parameters. Such a point was discussed in Ref.
\cite{Zhao:2013xha} for the $NN^*(1440)$ interaction.

This paper is organized as follows. After introduction, we present
the deduction of the OPE potentials in Sec. \ref{sec2}. In Sec.
\ref{sec3}, we present the corresponding numerical results. We
summarize our results in Sec. \ref{sec4}.

\section{Deduction of the effective potentials}\label{sec2}

In this section, we will derive the one-pion-exchange interaction
potential between the charmed mesons in the $T$ doublet and their
antiparticles. These charmed-anticharmed meson pairs can form three
$S$-wave systems $D_1\bar{D}_1$, $D_1\bar{D}_2^*$ ($D_2^*\bar{D}_1$)
and $D_2^*\bar{D}_2^*$, which are simply labeled $T\bar{T}$. The
P-parity of the systems are all even while the C-parity of the
second system can be even or odd. In studying the bound state
problem, we need the masses of the charmed mesons in the $T$ doublet
which are taken from the particle data group \cite{Agashe:2014kda}:
$M_{D_1}=2422.35$ MeV, $M_{D_2}=2463.50$ MeV.

\subsection{The wave functions}\label{subsecflavor}

To construct the wave function of a $T\bar{T}$ system with a given
C-parity, the convention for the C-parity transformation of a
charmed meson should be addressed. It is required to be consistent
with the convention in the Lagrangian (Eq. (\ref{Ctrans})). In
addition, for the $D_1\bar{D}_2^*$ ($D_2^*\bar{D}_1$) system, a
factor coming from the exchange of the two bosons appears. Along
with the procedure in Ref. \cite{Zhu:2013sca}, we construct the
flavor wave functions as follows. It is enough for us to consider
only the neutral states because of the isospin invariance. For the
$D_1\bar{D}_1$ and $D_2^*\bar{D}_2^*$ systems, one has
\begin{eqnarray}
|X_{T\bar{T}}^{0}[J]\rangle = \frac{1}{\sqrt{2}}
            \left(|T^0\bar{T}^{0}\rangle -x|T^+{T}^{-}\rangle\right),\label{wf1}
\end{eqnarray}
where $x=1$ ($-1$) for the isovector (isoscalar) case and $J$ is the
total angular momentum. For the $D_1\bar{D}_2^*(D_2^*\bar{D}_1)$
system, it is necessary to construct a G-parity eigenfunction in
order to get the correct potential for a given C-parity
\cite{Liu:2013rxa}. With the convention $D_1\leftrightarrow \bar{D}_1, D_2^*\leftrightarrow -\bar{D}_2^*$ under the C-parity transformation, one finds
\begin{eqnarray}
|X_{D_1\bar{D}_2^*}^{0}[J]\rangle &=&
       \frac{1}{{2}\sqrt{2}}
\left(\Big[|D_1^0\bar{D}_2^{*0}\rangle
       +(-1)^{J-3}|\bar{D}_2^{*0}D_1^0\rangle\Big]\right.\nonumber\\
&&-x\Big[|D_1^+{D}_2^{*-}\rangle+(-1)^{J-3}|{D}_2^{*-}D_1^+\rangle\Big]\nonumber\\
&&-c\Big[|\bar{D}_1^{0}D_2^{*0}\rangle+(-1)^{J-3}|D_2^{*0}\bar{D}_1^{0}\rangle\Big]\nonumber\\
&&\left.+xc\Big[|{D}_1^{-}D_2^{*+}\rangle+(-1)^{J-3}|D_2^{*+}{D}_1^{-}\rangle\Big]\right),\label{wf2}
\end{eqnarray}
where $c=+$ or $-$ is the C-parity of the system. In the following discussions, we simply use the notation $D_1\bar{D}_2^*$
to denote the state $D_1\bar{D}_2^*+D_2^*\bar{D}_1$ or
$D_1\bar{D}_2^*-D_2^*\bar{D}_1$.

The meson-antimeson molecular system is similar to the well-known
deuteron \cite{Tornqvist:1993vu,Tornqvist:1993ng} and the tensor
force might be important. We also include the coupled channel
effects due to the $D$-wave interaction between the constituents.
The allowed quantum numbers and possible coupled channels for the
systems are
\begin{eqnarray}
&&D_1\bar{D}_1
\left\{\begin{array}{cccc}  J^{PC}=0^{++}:  &|^1S_0\rangle,    &|^5D_0\rangle,\\
                            J^{PC}=1^{+-}:  &|^3S_1\rangle,    &|^3D_1\rangle,\\
                            J^{PC}=2^{++}:  &|^5S_2\rangle,    &|^1D_2\rangle,   &|^5D_2\rangle,
\end{array}\right.\label{spin1}\\
&&D_1\bar{D}_2^* \left\{\begin{array}{ccccc}
J^{PC}=1^{+\pm}:     &|^3S_1\rangle,    &|^3D_1\rangle,   &|^5D_1\rangle,   &|^7D_1\rangle,\\
J^{PC}=2^{+\pm}:     &|^5S_2\rangle,    &|^3D_2\rangle,   &|^5D_2\rangle,   &|^7D_2\rangle,\\
J^{PC}=3^{+\pm}:     &|^7S_3\rangle,    &|^3D_3\rangle,
&|^5D_3\rangle,  &|^7D_3\rangle,
\end{array}\right.\label{spin2}
\end{eqnarray}
\begin{eqnarray}
&&D_2^*\bar{D}_2^* \left\{\begin{array}{ccccc}
J^{PC}=0^{++}:   &|^1S_0\rangle,    &|^5D_0\rangle,\\
J^{PC}=1^{+-}:   &|^3S_1\rangle,    &|^3D_1\rangle,   &|^7D_1\rangle,\\
J^{PC}=2^{++}:   &|^5S_2\rangle,    &|^1D_2\rangle,   &|^5D_2\rangle,   &|^9D_2\rangle,\\
J^{PC}=3^{+-}:   &|^7S_3\rangle,    &|^3D_3\rangle,   &|^7D_3\rangle,\\
J^{PC}=4^{++}:   &|^9S_4\rangle,    &|^5D_4\rangle,
&|{{}^9D_4}\rangle,
\end{array}\right.\label{spin3}
\end{eqnarray}
where the notation $|^{2S+1}L_J\rangle$ is used. Contributions from
the higher partial wave channels are significantly suppressed and we
do not include such channels here. The P-parity is always +1. The
C-parity for the $D_1\bar{D}_1$ and $D_2^*\bar{D}_2^*$ systems is
determined with $C=(-1)^{L+S}$ while the C-parity for the
$D_1\bar{D}_2^*$ system is not constrained by symmetry. Since the
orbital angular momentum $L=0, 2$, one observes that the total spin
$S$ of the system is always odd (even) for the $D_1\bar{D}_1$ and
$D_2^*\bar{D}_2^*$ states with negative (positive) C-parity while
both odd and even $S$ are allowed for the $D_1\bar{D}_2^*$ states.

The molecular system composed of two mesons $AB$ is defined as
\begin{eqnarray}
|AB(^{2S+1}L_J)\rangle=\sum_{m_A,m_B,m_S,m_L}C_{s_A,m_A;s_B,m_B}^{S,m_S}C_{S,m_S;L,m_L}^{J,M}
\epsilon^{m_A}\epsilon^{m_B}|Y_{L,m_L}\rangle.
\end{eqnarray}
Here, $C_{s_A,m_A;s_B,m_B}^{S,m_S}$ and $C_{S,m_S;L,m_L}^{J,M}$
denote the CG coefficients, $|Y_{L,m_L}\rangle$ is the spherical
harmonic function, and $\epsilon^m$ is the polarization vector
$\varepsilon^m$ for the axial-vector meson or the polarization
tensor $\xi^m$ for the tensor meson. In the static limit, the
special expressions for the polarization vector are
$\varepsilon^{\pm1}= \frac{1}{\sqrt{2}}\left(0,\pm1,i,0\right)$ and
$\varepsilon^{0}= \left(0,0,0,-1\right)$. The polarization tensor
$\xi^{m}$ can be constructed from the polarization vectors, $\xi^{m}
=\sum_{m_1,m_2}C_{1,m_1;1,m_2}^{2,m}\varepsilon^{m_1}\varepsilon^{m_2}$
\cite{Cheng:2010yd}. One may use each z-component
$(M=0,\pm1,\cdots,\pm J)$ to derive the potential.

\subsection{The effective Lagrangian}\label{subsec2}

We will study the meson-antimeson interaction in a one-pion-exchange
(OPE) model. The effective potential is derived using the effective
Lagrangian with the heavy quark symmetry and chiral symmetry, which
can be written in a compact form \cite{Ding:2008gr},
\begin{eqnarray}\label{lagd1d2}
\mathcal{L}_{\mathbb{P}}&=& ik\langle T^{(Q)\mu}_b{\rlap\slash
A}_{ba}\gamma_5\bar{T}^{(Q)}_{a\mu}\rangle +ik\langle
\bar{T_a}^{(\bar{Q})\mu}\rlap\slash
A_{ab}\gamma_5T_{b\mu}^{(\bar{Q})}\rangle.
\end{eqnarray}
Here, the axial field $A_\mu$ is constructed with the pion field
\begin{eqnarray}
A_{\mu} &=& \frac{1}{2}(\zeta^{\dag}\partial_{\mu}\zeta-\zeta\partial_{\mu}\zeta^{\dag}),\nonumber\\
\zeta&=&\exp(i\mathbb{P}/f_{\pi}),\nonumber\\
\mathbb{P} &=&\left(\begin{array}{cc}
\frac{\pi^0}{\sqrt{2}} &\pi^+  \nonumber\\
\pi^- &-\frac{\pi^0}{\sqrt{2}}
\end{array}\right),
\end{eqnarray}
where $f_{\pi}=132$ MeV is the pion decay constant. The $T$ doublet
now is expressed with a form of the superfield
\begin{eqnarray*}\label{eqTQ}
T_a^{(Q)\mu} &=& \frac{1+\rlap\slash
v}{2}[P_{2a}^{*(Q)\mu\nu}\gamma_{\nu}
                   -\sqrt{\frac{3}{2}}P_{1a\nu}^{(Q)}\gamma_5(g^{\mu\nu}-\frac{1}{3}\gamma^{\nu}
                 (\gamma^{\mu}-v^{\mu}))],\nonumber\\
P_{2a}^{*(Q)}&=&(D_2^{*0},D_2^{*+}),\qquad\qquad
P_1^{(Q)}=(D_1^0,D_1^+),
\end{eqnarray*}
where the four velocity $v=(1, \vec{0})$ in the static
approximation. The normalization for the involved mesons are
$\langle
0|P_1^{\mu}|Q\bar{q}(1^+)\rangle=\varepsilon^{\mu}\sqrt{M_{P_1}}$
and $\langle
0|P_2^{*\mu\nu}|Q\bar{q}(2^+)\rangle=\xi^{\mu\nu}\sqrt{M_{P_2^*}}$.
The antimeson doublet containing a heavy anti-quark $\bar{Q}$ is
obtained from the charge conjugate operation,
\begin{eqnarray}\label{eqTQbar}
T_a^{(\bar{Q})\mu} &=& C\left(\mathcal{C}T_a^{(Q)\mu}\mathcal{C}^{-1}\right)^TC^{-1}\nonumber\\
                   &=& [P_{2a}^{*(\bar{Q})\mu\nu}\gamma_{\nu}
                 -\sqrt{\frac{3}{2}}P_{1a\nu}^{(\bar{Q})}\gamma_5(g^{\mu\nu}-\frac{1}{3}\gamma^{\nu}
                 (\gamma^{\mu}-v^{\mu})\gamma^{\nu})]\frac{1-\rlap\slash v}{2},\nonumber
\end{eqnarray}
where the transpose is for the $\gamma$ matrices, $\mathcal{C}$ is
the charge conjugate operator and the matrix $C=i\gamma^2\gamma^0$.
The convention for the C-parity transformation for meson and
antimeson fields is
\begin{eqnarray}\label{Ctrans}
P_{2a}^{*(\bar{Q})\mu\nu}=-\mathcal{C}P_{2a}^{*({Q})\mu\nu}\mathcal{C}^{-1},
\quad\quad\quad
P_{1a\nu}^{(\bar{Q})}=\mathcal{C}P_{1a\nu}^{({Q})}\mathcal{C}^{-1},
\end{eqnarray}
and the hermitian conjugate fields are defined by
\begin{eqnarray}
\bar{T}_a^{(Q)\mu}=\gamma_0T_a^{(Q)\mu\dag}\gamma_0,~~~~~~~~~~~~~~~
\bar{T}_{a\mu}^{(\bar{Q})}=\gamma_0T_{a\mu}^{(\bar{Q})\dag}\gamma_0.
\end{eqnarray}
In the compact Lagrangian, $\langle...\rangle$ denotes trace in the
spin and flavor space.

Expanding Eq. (\ref{lagd1d2}), one obtains the interaction terms of
the $D_1(D_2^*)$ and $\bar{D}_1(\bar{D}_2^*)$ mesons with the pion,
 \begin{eqnarray}
 \mathcal{L}_{D_1D_1\mathbb{P}} &=&
            i\frac{5k}{3f_{\pi}}v^{\mu}\varepsilon_{\mu\nu\alpha\beta}
            D_{1b}^{\nu}D_{1a}^{\beta\dag}\partial^{\alpha}\mathbb{P}_{ba}\nonumber\\
            &&-i\frac{5k}{3f_{\pi}}v^{\mu}\varepsilon_{\mu\nu\alpha\beta}
            \bar{D}_{1a}^{\nu\dag}\bar{D}_{1b}^{\beta}\partial^{\alpha}\mathbb{P}_{ab},\nonumber\\
 %%%%%%%%%%%%%%%%%%%%%%%%%%%%%%%%%%%%
 \mathcal{L}_{D_2^*D_2^*\mathbb{P}} &=&   -2i\frac{k}{f_{\pi}}
            v^{\lambda}\varepsilon_{\lambda\nu\alpha\sigma}
            D_{2b}^{*\mu\nu}D_{2a\mu}^{*\sigma\dag}\partial^{\alpha}\mathbb{P}_{ba}\nonumber\\
            &&-2i\frac{k}{f_{\pi}}v^{\lambda}\varepsilon_{\lambda\nu\alpha\sigma}
            \bar{D}_{2a\mu}^{*\sigma\dag}\bar{D}_{2b}^{*\mu\nu}\partial^{\alpha}\mathbb{P}_{ab},\nonumber\\
 %%%%%%%%%%%%%%%%%%%%%%%%%%%%%%%%%%%%%%%
 \mathcal{L}_{D_1D_2^*\mathbb{P}} &=& -\sqrt{\frac{2}{3}}\frac{k}{f_{\pi}}
           \left(D_{2b}^{*\mu\nu}D_{1a\mu}^{\dag}+D_{1b\mu}D_{2a}^{*\mu\nu\dag}\right)
           \partial_{\nu}\mathbb{P}_{ba}\nonumber\\
           &&+\sqrt{\frac{2}{3}}\frac{k}{f_{\pi}}
           \left(\bar{D}_{2a}^{*\mu\nu\dag}\bar{D}_{1b\mu}
           +\bar{D}_{1a\mu}^{\dag}\bar{D}_{2b}^{*\mu\nu}\right)\partial_{\nu}\mathbb{P}_{ab}.
 \end{eqnarray}

The lowest charmed meson doublet is $H=(0^-,1^-)$. The coupling
constant $g$ for the $H-H-\pi$ interaction can be extracted from the
strong decay $D^*\to D\pi$: $g=0.59\pm 0.07\pm 0.01$
\cite{Isola:2003fh}. In this work, the coupling constant $k$ is
taken as the same value of $g$
\cite{Liu:2008xz,Chen:2014mwa,Falk:1992cx}.

\subsection{The OPE effective potential}

With the above effective Lagrangians, we can write down the
amplitudes for the t-channel scattering processes, which are related
to the required potentials. The involved processes are
$D_1\bar{D}_1\rightarrow D_1\bar{D}_1$, $D_1\bar{D}_2^*\rightarrow
D_1\bar{D}_2^*$, $D_1\bar{D}_2^*\rightarrow D_2^*\bar{D}_1$,
$D_2^*\bar{D}_1\rightarrow D_1\bar{D}_2^*$,
$D_2^*\bar{D}_1\rightarrow D_2^*\bar{D}_1$, and
$D_2^*\bar{D}_2^*\rightarrow D_2^*\bar{D}_2^*$. Generally, the
effective potential in the momentum space is related to the obtained
scattering amplitude through the Breit approximation. Here, we take
the process $D_1\bar{D}_1\rightarrow D_1\bar{D}_1$ as an example.
The relation between the effective potential $\mathcal{V}$ and
scattering amplitude $\mathcal{M}$ is
\begin{eqnarray}\label{Breit}
\mathcal{V}\left[{D_1\bar{D}_1\rightarrow
D_1\bar{D}_1}\right](\vec{q}) =
          -\frac{\mathcal{M}(D_1\bar{D}_1\rightarrow D_1\bar{D}_1)}
          {\sqrt{\prod_i2M_i\prod_f2M_f}},
\end{eqnarray}
where $M_i$ and $M_f$ denote the masses of the initial and final
states, respectively. We calculate the scattering amplitude
$\mathcal{M}(D_1\bar{D}_1\rightarrow D_1\bar{D}_1)$ in the
one-pion-exchange approximation. In addition, one obtains the
effective potential in the coordinate space by performing the
Fourier transformation,
\begin{eqnarray}\label{form fact}
&&\mathcal{V}\left[{D_1\bar{D}_1\rightarrow D_1\bar{D}_1}\right](\vec{r})\nonumber\\
 &=&\int\frac{d^3\vec{q}}{(2\pi)^3}e^{i\vec{q}\cdot\vec(r)}
\mathcal{F}^2(q^2,m_E^2)\mathcal{V}\left[{D_1\bar{D}_1\rightarrow
D_1\bar{D}_1}\right](\vec{q}),\nonumber
\end{eqnarray}
where $\mathcal{F}(q^2,m_E^2)=(\Lambda^2-m_E^2)/(\Lambda^2-q^2)$ is
the monopole type form factor with $m_E$ the mass of the exchanged
meson. The monopole form factor is introduced at each interaction
vertex to compensate the off shell effect of the exchanged meson and
describe the structure effect of the vertex. In practice, the
phenomenological cutoff $\Lambda$ is around one to several GeV.

Following the above procedure, we obtain the OPE potentials for the
$D_1\bar{D}_1$, $D_1\bar{D}_2^*$ and $D_2^*\bar{D}_2^*$ systems. The
form of the final potentials for these systems may be summarized
into a compact form
\begin{eqnarray}
\mathcal{V}_{D_1\bar{D}_1}(I,J,r) &=& \mathcal{G}(I)\times\mathcal{V}_1,\label{totalpential1}\\
\mathcal{V}_{D_1\bar{D}_2^*}(I,J,r) &=&
                 \frac{1}{2}\mathcal{G}(I)\times
                 \left(\mathcal{V}_5+\mathcal{V}_8
                 -c(-1)^{J-3}\left(\mathcal{V}_6+\mathcal{V}_{11}\right)\right),
                 \label{totalpential2}\nonumber\\\\
\mathcal{V}_{D_2^*\bar{D}_2^*}(I,J,r) &=&
\mathcal{G}(I)\times\mathcal{V}_{10}.\label{totalpential3}
\end{eqnarray}
Here, $I$ is the isospin of the state and $\mathcal{G}(I)$ is
defined as an isospin factor, with the value $\mathcal{G}(0)=3/2$
and $\mathcal{G}(1)=-1/2$. One gets these potentials by sandwiching
the above $(I,J)$-independent potentials between the states in Eqs.
(\ref{wf1}) and (\ref{wf2}).

We take the $D_1\bar{D}_1$ system as an example to illustrate the
structure of the potentials. Now
\begin{eqnarray}
\mathcal{V}_{D_1\bar{D}_1}(I,J,r) &=& \langle X^0_{D_1\bar{D}_1}[J]|\mathcal{V}\left[{D_1\bar{D}_1\rightarrow D_1\bar{D}_1}\right](\vec{r})|X^0_{D_1\bar{D}_1}[J]\rangle\nonumber\\
&\equiv&\langle X^0_{D_1\bar{D}_1}[J]|\hat{\mathcal{V}}_1|X^0_{D_1\bar{D}_1}[J]\rangle\nonumber\\
&=&\mathcal{G}(I)\times\mathcal{V}_1,\label{D1D1pot}
\end{eqnarray}
where $\hat{\mathcal{V}}_1$ has the form
$\hat{\mathcal{V}}_1=const\times[\mathcal{E}_1f(r)+\mathcal{S}_1g(r)]$.
Here, $\mathcal{E}_1$ ($\mathcal{S}_1$) is the spin-spin (tensor)
operator and $f(r)$ ($g(r)$) is a potential function depending on
the cutoff $\Lambda$, the pion mass $m_\pi$, and the radial
coordinate $r$. Obviously, $\mathcal{V}_1$ is a $n\times n$ matrix,
where $n$ is the number of the coupled channels in Eq. (\ref{spin1})
which depends on the total angular momentum $J$. The explicit matrix
elements for the operator $\mathcal{E}_1$ are
\begin{eqnarray}
\langle\mathcal{E}_1[J=0]\rangle &=& \left(\begin{array}{cc}
\langle{}^1S_0|\mathcal{E}_1|{}^1S_0\rangle    &\langle{}^1S_0|\mathcal{E}_1|{}^5D_0\rangle\\
\langle{}^5D_0|\mathcal{E}_1|{}^1S_0\rangle
&\langle{}^5D_0|\mathcal{E}_1|{}^5D_0\rangle
\end{array}\right)=\left(\begin{array}{cc}2  &0\\
0  &-1\end{array}\right),\nonumber\\
\langle\mathcal{E}_1[J=1]\rangle &=& \left(\begin{array}{cc}
\langle{}^3S_1|\mathcal{E}_1|{}^3S_1\rangle    &\langle{}^3S_1|\mathcal{E}_1|{}^3D_1\rangle\\
\langle{}^3D_1|\mathcal{E}_1|{}^3S_1\rangle
&\langle{}^3D_1|\mathcal{E}_1|{}^3D_1\rangle
\end{array}\right)
=\left(\begin{array}{cc}1   &0 \\
0   &1\end{array}\right),\nonumber\\
\langle\mathcal{E}_1[J=2]\rangle &=& \left(\begin{array}{ccc}
\langle{}^5S_2|\mathcal{E}_1|{}^5S_2\rangle
&\langle{}^5S_2|\mathcal{E}_1|{}^1D_2\rangle
&\langle{}^5S_2|\mathcal{E}_1|{}^5D_2\rangle\\
\langle{}^1D_2|\mathcal{E}_1|{}^5S_2\rangle
&\langle{}^1D_2|\mathcal{E}_1|{}^1D_2\rangle
&\langle{}^1D_2|\mathcal{E}_1|{}^5D_2\rangle\\
\langle{}^5D_2|\mathcal{E}_1|{}^5S_2\rangle
&\langle{}^5D_2|\mathcal{E}_1|{}^1D_2\rangle
&\langle{}^5D_2|\mathcal{E}_1|{}^5D_2\rangle\\
\end{array}\right)\nonumber\\
&=&\left(\begin{array}{ccc}-1   &0    &0\\
                                      0    &2     &0\\
                                      0    &0     &-1\end{array}\right),
\end{eqnarray}
and those for $\mathcal{S}_1$ are
\begin{eqnarray}
\langle\mathcal{S}_1[J=0]\rangle &=&
   {\left(\begin{array}{cc}0&\sqrt{2}\\\sqrt{2}  &2\end{array}\right)_{J=0}},\nonumber\\
\langle\mathcal{S}_1[J=1]\rangle &=&
   {\left(\begin{array}{cc}0&-\sqrt{2}\\-\sqrt{2}&1\end{array}\right)_{J=1}},\nonumber\\
\langle\mathcal{S}_1[J=2]\rangle &=&
   {\left(\begin{array}{ccc}0&\sqrt{\frac{2}{5}}&-\sqrt{\frac{14}{5}}\\\sqrt{\frac{2}{5}}&0&-\frac{2}{\sqrt{7}}\\
                 -\sqrt{\frac{14}{5}}  &-\frac{2}{\sqrt{7}}    &-\frac{3}{7}\end{array}\right)_{J=2}}.
\end{eqnarray}

When the quantum numbers are the same, the transitions between the
states $D_1\bar{D}_1$, $D_1\bar{D}_2^*$, and $D_2^*\bar{D}_2^*$ may
happen. We also consider such coupled channel effects in this paper.
The involved transitions include $D_1\bar{D}_1\leftrightarrow
D_1\bar{D}_2^*$, $D_1\bar{D}_1\leftrightarrow D_2^*\bar{D}_1$,
$D_1\bar{D}_1\leftrightarrow D_2^*\bar{D}_2^*$,
$D_1\bar{D}_2^*\leftrightarrow D_2^*\bar{D}_2^*$, and
$D_2^*\bar{D}_1\leftrightarrow D_2^*\bar{D}_2^*$. Relevant
transition potentials are $\hat{\mathcal{V}}_{2,3,4,7,9}$. We will
show the form of the final potentials in the next section. We
collect all the defined subpotentials
$\hat{\mathcal{V}}_k(k=1,2,\cdots,11)$ in Eqs.
(\ref{poten1})-(\ref{poten11}) in Appendix \ref{app01}. The related
matrix elements are given there in Tables \ref{ss} and \ref{tensor}.

%%%%%%%%%%%%%%%%%%%%%%%%%%%%%%%%%%%%%%%%%%%%%%%%%%%
\section{Numerical results}\label{sec3}
%%%%%%%%%%%%%%%%%%%%%%%%%%%%%%%%%%%%%%%%%%%%%%%%%%%

With the effective potentials
(\ref{totalpential1})-(\ref{totalpential3}), we can search for the
bound state solutions by solving the coupled-channel Schr\"{o}dinger
equation, {\begin{eqnarray}
\frac{1}{2\mu}\left(-\nabla^2+\frac{l(l+1)}{r^2}\right)
\psi(r)+(V(r)-E)\psi(r)=0,\label{sch}
\end{eqnarray}
where $\nabla^2 = \frac{1}{r^2}\frac{d}{d r}r^2\frac{d}{d r}$,
$V(r)=\mathcal{V}_{D_1\bar{D}_1}(I,J,r)$,
$\mathcal{V}_{D_1\bar{D}_2^*}(I,J,r)$, or
$\mathcal{V}_{D_2^*\bar{D}_2^*}(I,J,r)$, and $\mu$ is the reduced
mass of the system. The wave function $\psi(r)$ is a column matrix
depending on $V(r)$.

Then we adopt the FESSDE program
\cite{Abrashkevichn:1995cj,Abrashkevichn:1998cj} to calculate the
energy eigenvalue $E$ in units of MeV. If a bound state solution
exists, the binding energy $E$ is defined as the mass difference
$M_{system}-M_{constituents}$. We further calculate the
corresponding root-mean-square (RMS) radius $r_{RMS}$ in units of
fm. $E$ and $r_{RMS}$ depend on the cutoff parameter $\Lambda$ which
varies from 0.5 to 5 GeV.

\subsection{The pure $D_1\bar{D}_1$, $D_1\bar{D}_2^*$ and $D_2^*\bar{D}_2^*$ systems}\label{sec3sub01}

In this subsection, we present three sets of numerical results for
the pure $D_1\bar{D}_1$, $D_1\bar{D}_2^*$ and $D_2^*\bar{D}_2^*$
systems in Tables \ref{pure1} and \ref{pure2}. There are fourteen
isoscalar and four isovector bound state solutions in the $T\bar{T}$
systems. The $\Lambda$ dependence of the obtained binding energy is
shown in FIG. \ref{pure}. The results are very sensitive to the cutoff parameter, which is a common feature in these types of investigations \cite{Liu:2008fh,Liu:2008tn,Sun:2011uh,Sun:2012zzd,
Liu:2008xz,Hu:2010fg,Shen:2010ky,Yang:2011wz,Sun:2012sy,
Li:2012ss,Chen:2014mwa,Zhu:2013sca,Ding:2008gr,Zhang:2006ix,
Chen:2013sba,Thomas:2008ja,Lee:2009hy}. At present, we cannot determine its value for the present systems from experimental observables. One should be cautious that the binding solution for the meson-antimeson systems are possible only when the cutoff falls into a reasonable range. A repulsive potential or attractive potential with an inappropriate cutoff does not result in bound states.

\renewcommand{\arraystretch}{1.5}
\begin{table}[!hbtp]
\caption{The bound state solutions (the binding energy $E$ and the
root-mean-square radius $r_{RMS}$) for the pure $D_1\bar{D}_1$ and
$D_2^*\bar{D}_2^*$ systems. $E$, $r_{RMS}$, and $\Lambda$ are in
units of MeV, fm, and GeV, respectively. The notation $\cdots$ means
that no binding solution is found.}\label{pure1}
\begin{tabular}{l|cccc|cccc}
\toprule[1pt] \toprule[1pt] Systems    &$I^G(J^{PC})$    &$\Lambda$
&$E$    &$r_{RMS}$
           &$I^G(J^{PC})$    &$\Lambda$   &$E$    &$r_{RMS}$
           \\\midrule[1pt]
$D_1\bar{D}_1$
          &$0^+(0^{++})$     &1.22     &-0.42     &4.21
          &$1^-(0^{++})$     &\ldots&\ldots&\ldots\\
          &                  &1.32     &-3.41     &1.67
          &                  &\ldots&\ldots&\ldots\\
          &                  &1.42     &-9.64     &1.06
          &                  &\ldots&\ldots&\ldots\\
          &$0^-(1^{+-})$     &1.96     &-1.34     &2.59
          &$1^+(1^{+-})$     &\ldots&\ldots&\ldots\\
          &                  &2.06    &-4.35    &1.50
          &                  &\ldots&\ldots&\ldots\\
          &                  &2.16    &-9.38    &1.06
          &                  &\ldots&\ldots&\ldots\\
          &$0^+(2^{++})$     &3.49    &-1.02    &3.14
          &$1^-(2^{++})$     &4.75    &-1.43    &2.44\\
          &                  &3.59    &-2.27    &2.20
          &                  &4.85    &-3.68    &1.55\\
          &                  &3.69    &-4.12    &1.70
          &                  &4.95    &-7.08    &1.14\\\hline
$D_2^*\bar{D}_2^*$
          &$0^+(0^{++})$     &1.10   &-0.33     &4.44
          &$1^-(0^{++})$     &\ldots&\ldots&\ldots\\
          &                  &1.20   &-3.21     &1.74
          &      &\ldots&\ldots&\ldots\\
          &                  &1.30   &-9.46     &1.08
          &      &\ldots&\ldots&\ldots\\
          &$0^-(1^{+-})$     &1.22   &-0.07     &5.76
          &$1^+(1^{+-})$     &\ldots&\ldots&\ldots\\
          &                  &1.32   &-2.04     &2.12
          &      &\ldots&\ldots&\ldots\\
          &                  &1.42   &-6.74     &1.25
          &      &\ldots&\ldots&\ldots\\
          &$0^+(2^{++})$     &1.74   &-0.42     &4.16
          &$1^-(2^{++})$     &\ldots&\ldots&\ldots\\
          &                  &1.84   &-2.56     &1.91
          &                  &\ldots&\ldots&\ldots\\
          &                  &1.94   &-6.80     &1.23
          &                  &\ldots&\ldots&\ldots\\
          &$0^-(3^{+-})$     &2.70   &-0.24     &4.97
          &$1^+(3^{+-})$     &\ldots&\ldots&\ldots\\
          &                  &2.90   &-3.08     &1.85
          &                  &\ldots&\ldots&\ldots\\
          &                  &3.10   &-9.94     &1.11
          &                  &\ldots&\ldots&\ldots\\
          &$0^+(4^{++})$     &3.90   &-0.29     &4.86
          &$1^-(4^{++})$     &3.58   &-0.01     &6.15\\
          &                  &4.20   &-2.61     &2.07
          &                  &3.78   &-3.26     &1.64\\
          &                  &4.50   &-8.23     &1.26
          &                  &3.98   &-12.09    &0.89\\
\bottomrule[1pt] \bottomrule[1pt]
\end{tabular}
\end{table}

\renewcommand{\arraystretch}{1.5}
\begin{table}[!hbtp]
\caption{The bound state solutions (the binding energy $E$ and the
root-mean-square radius $r_{RMS}$) for the pure $D_1\bar{D}_2^*$
system. $E$, $r_{RMS}$, and $\Lambda$ are in units of MeV, fm, and
GeV, respectively. The notation $\cdots$ means that no binding
solution is found.}\label{pure2}
\begin{tabular}{l|cccc|cccc}
\toprule[1pt] \toprule[1pt] Systems    &$I^G(J^{PC})$    &$\Lambda$
&$E$    &$r_{RMS}$
           &$I^G(J^{PC})$    &$\Lambda$   &$E$    &$r_{RMS}$
           \\\midrule[1pt]
$D_1\bar{D}_2^*$
          &$0^-(1^{+-})$     &1.32    &-0.17    &5.11
          &$1^+(1^{+-})$     &\ldots&\ldots&\ldots\\
          &                  &1.42   &-2.62     &1.84
          &      &\ldots&\ldots&\ldots\\
          &                  &1.52   &-8.13     &1.11
          &      &\ldots&\ldots&\ldots\\
          &$0^-(2^{+-})$     &3.36   &-0.53     &3.75
          &$1^+(2^{+-})$     &\ldots&\ldots&\ldots\\
          &                  &3.46   &-2.25     &1.97
          &      &\ldots&\ldots&\ldots\\
          &                  &3.56   &-5.25     &1.32
          &      &\ldots&\ldots&\ldots\\
          &$0^-(3^{+-})$     &3.03   &-1.17     &2.96
          &$1^+(3^{+-})$     &4.71   &-0.82     &3.14\\
          &                  &3.13   &-2.70     &2.06
          &                  &4.81   &-2.56     &1.84\\
          &                  &3.23   &-5.04     &1.57
          &                  &4.91   &-5.35     &1.30\\\cline{2-9}
          &$0^+(1^{++})$     &1.34   &-0.31     &4.48
          &$1^-(1^{++})$     &\ldots&\ldots&\ldots\\
          &                  &1.44   &-2.94     &1.77
          &      &\ldots&\ldots&\ldots\\
          &                  &1.54   &-8.50     &1.10
          &      &\ldots&\ldots&\ldots\\
          &$0^+(2^{++})$     &2.30   &-0.37     &4.29
          &$1^-(2^{++})$     &\ldots&\ldots&\ldots\\
          &                  &2.40   &-2.08     &2.08
          &      &\ldots&\ldots&\ldots\\
          &                  &2.50   &-5.33     &1.36
          &      &\ldots&\ldots&\ldots\\
          &$0^+(3^{++})$     &3.91   &-0.25     &5.06
          &$1^-(3^{++})$     &3.98   &-1.04     &2.79\\
          &                  &4.21   &-3.48     &1.82
          &                  &4.08   &-3.35     &1.60\\
          &                  &4.45   &-9.89     &1.16
          &                  &4.18   &-7.08     &1.12\\
\bottomrule[1pt] \bottomrule[1pt]
\end{tabular}
\end{table}

\begin{figure*}[!htbp]
  \centering
\includegraphics[width=7.0in]{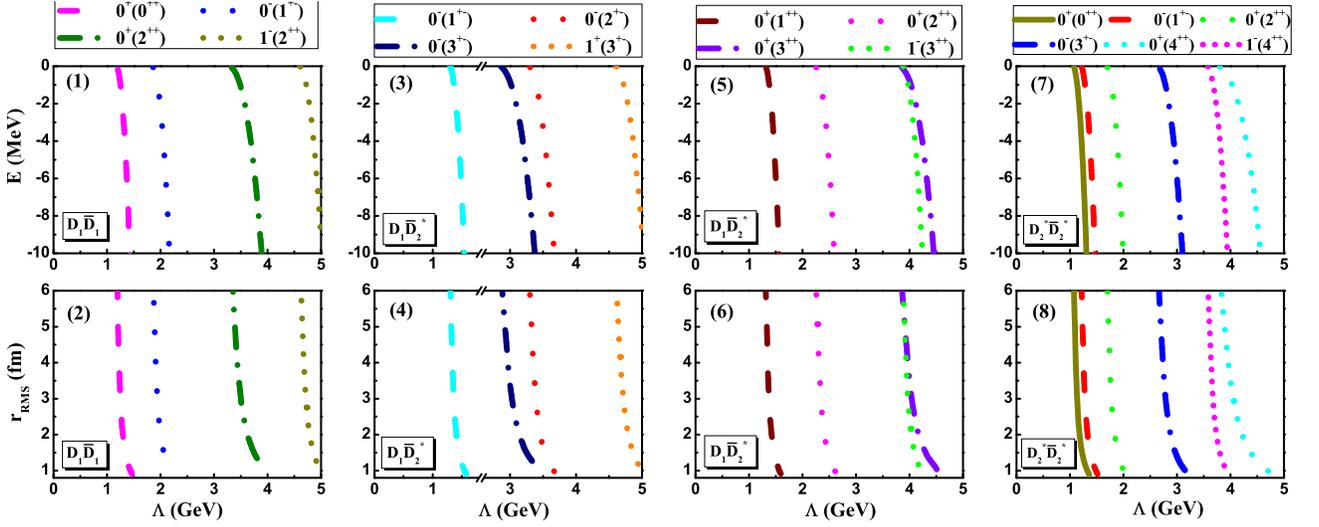}
\caption{(color online). The $\Lambda$ dependence of the bound state
solutions (the binding energy $E$ and the root-mean-square radius
$r_{RMS}$) for the pure $D_1\bar{D}_1$, $D_1\bar{D}_2^*$ and
$D_2^*\bar{D}_2^*$ systems with all the possible
configurations.}\label{pure}
\end{figure*}

If we compare all the bound state solutions system by system, we
notice an inequality
$\Lambda_{J=0}^{I=0}(D_1\bar{D}_1)<\Lambda_{J=1}^{I=0}(D_1\bar{D}_1)<\Lambda_{J=2}^{I=0}(D_1\bar{D}_1)$
if the binding energies are fixed at roughly the same value. To some
extent, the existence possibility of one molecule is related to the
value of the cutoff $\Lambda$. We notice
$P(0^{+}(0^{++}))>P(0^{-}(1^{+-}))>P(0^{+}(2^{++}))$, where
$P(0^{+}(0^{++}))$ denotes the existence possibility for the
$D_1\bar{D}_1$ with $I^G(J^{PC})=0^{+}(0^{++})$. In other words, the
$D_1\bar{D}_1$ with $I^G(J^{PC})=0^{+}(0^{++})$ is the most stable
loosely bound structure, followed by $0^{-}(1^{+-})$ and
$0^{+}(2^{++})$. This binding feature results certainly from the potentials. Since the main contribution comes from the S-wave interaction, the matrix elements for the operator ${\cal E}_1$ in the diagonal S-wave channels, $\langle ^1S_0|{\cal E}_1|^1S_0\rangle=2$, $\langle ^3S_1|{\cal E}_1|^3S_1\rangle=1$, and $\langle ^5S_2|{\cal E}_1|^5S_2\rangle=-1$, affect dominantly the attraction strength. Therefore, the $0^{++}$ state is the most easily bound. The isoscalar scalar $D_1\bar{D}_1$ system may be
easiest to detect in future experiments. Moreover, we also notice
the same trend in the isoscalar $D_1\bar{D}_2^*$ and isoscalar
$D_2^*\bar{D}_2^*$ cases.

In terms of the isovector states, there are only four binding
solutions, which are the $D_1\bar{D}_1$ system with $1^-(2^{++})$,
$D_1\bar{D}_2^*$ with $1^+(3^{+-}),1^-(3^{++})$ and
$D_2^*\bar{D}_2^*$ with $1^-(4^{++})$. All these $T\bar{T}$ systems
could be easily observed if they do exist since the charged state is
easy to identify experimentally. Moreover, their total angular
momentum $J$ always takes the largest value for these systems.
The reason is that the sign for the potential in the isovector case is opposite to that in the isoscalar case. A smaller matrix element for the operator ${\cal E}_1$ in the diagonal S-wave channel leads to a stronger attraction. From the matrix elements in Table \ref{ss}, the smallest value usually corresponds to the largest $J$. Compared to the isoscalar states, the values of the cutoff $\Lambda$ in the isovector case are all around 4 to 5 GeV. From the experience
of the deuteron ($\Lambda\sim 1$ GeV), this large cutoff means that
the attraction is not so strong. It is not difficult to understand this observation from the potentials,
where an isospin factor ${\cal G}(I)$ always exists. The magnitude of the factor (and thus the potential)
in the isovector case is smaller than that in the isoscalar case by a factor of three. As a result, one usually finds more binding solutions in the isoscalar case. % This seems to be a generic feature in the study for meson-antimeson bound state problem???...}
Further studies of the existence possibility for the isovector states are required.

When we ignore the channel couplings among $D_1\bar{D}_1$,
$D_1\bar{D}_2^*$, and $D_2^*\bar{D}_2^*$, the above results indicate
that the $D_1\bar{D}_1$ with $I^G(J^{PC})=0^+(0^{++})$,
$D_1\bar{D}_2^*$ with $0^-(1^{+-}),0^+(1^{++})$ and
$D_2^*\bar{D}_2^*$ with $0^+(0^{++}),0^-(1^{+-})$ can be filtrated
out as the molecule candidates if the reasonable cutoff in the OPE
model is around 1 GeV.

Searching for these prime molecular candidates is very interesting
in the future. In the following we also discuss their decay
behavior. We present all the allowed S-wave two-body decay modes.
The decay channels of the $D_1\bar{D}_1$ molecular state with
$0^+(0^{++})$ include $D\bar{D}$, $D^*\bar{D}^*$, $\eta_c(nS)\eta$
$(n=1,2)$ and $\psi(nS)\omega$ $(n=1,2,3)$. For the $D_1\bar{D}_2^*$
state with $0^-(1^{+-})$, its decay modes are $D\bar{D}^*$,
$D^*\bar{D}^*$, $\psi(nS)\eta$ $(n=1,2,3,4)$ and $\eta_c(nS)\omega$
$(n=1,2)$, while the other $D_1\bar{D}_2^*$ state with $0^+(1^{++})$
can decay into $D\bar{D}^*$ and $\psi(nS)\omega$ $(n=1,2,3)$.
$D\bar{D}$, $D^*\bar{D}^*$, $\eta_c(nS)\eta$ $(n=1,2)$ and
$\psi(nS)\omega$ $(n=1,2,3)$ would be the main decay channels for
the $D_2^*\bar{D}_2^*$ state with $0^+(0^{++})$. Additionally, the
$D_2^*\bar{D}_2^*$ state with $0^-(1^{+-})$ can decay to
$D\bar{D}^*$, $D^*\bar{D}^*$, $\psi(nS)\eta$ $(n=1,2,3,4)$ and
$\eta_c(nS)\omega$ $(n=1,2)$.

\subsection{The coupled system}\label{sec3sub02}

Since the charmed meson masses in the $T$ doublet $(D_1, D_2^*)$ are
very close, it is very necessary to consider the coupled channel
effects for various systems. In Table \ref{channel}, we present the
possible channels to be considered in this work. The states with
$J^{PC}=1^{++}$, $2^{+-}$, $3^{++}$, and $4^{++}$ appear only in
either $D_1\bar{D}_2^*$ or $D_2^*\bar{D}_2^*$ system, which have
been discussed in the above subsection.

Compared with Eqs. (\ref{spin1})-(\ref{spin3}), several $D$-wave
channels are omitted in Table \ref{channel} since their
contributions are small in our calculation. For illustration, we
present the contributions from each S-wave and D-wave channel to the
pure $D_1\bar{D}_2^*$ and $D_2^*\bar{D}_2^*$ systems for some cases
In Table \ref{poss}. It is obvious that the contributions from the
neglected channels such as $|^5D_1\rangle$, $|^7D_1\rangle$,
$|^7D_2\rangle$, $|^9D_2\rangle$, and $|^5D_3\rangle$ are small
(less than 0.6\%) in the $J=1$ case. In the coupled channel cases,
the contributions of the above neglected $D$-wave $D_1\bar{D}_2^*$
and $D_2^*\bar{D}_2^*$ channels are even smaller than those for the
pure systems. We have confirmed that the omission of these tiny
D-wave channels do not affect the main results after solving the
coupled Schrodinger equation.

Now, the binding energy is defined relative to the $D_1\bar{D}_1$
threshold. One has to add $(M_{D_2^*}-M_{D_1})$ to the kinetic term
in Eq. (\ref{sch}) for the $D_1\bar{D}_2^*$ channels and
$2(M_{D_2^*}-M_{D_1})$ for the $D_2^*\bar{D}_2^*$ channels in
solving the Schroding equation for the coupled system.

\renewcommand{\arraystretch}{1.5}
\begin{table*}[!hbtp]
\caption{Channels to be considered for the coupled
$D_1\bar{D}_1-D_1\bar{D}_2^*-D_2^*\bar{D}_2^*$
system.}\label{channel}
\begin{tabular}{c|ccccccccc}
  \toprule[1pt]\toprule[1pt]
%  &\multicolumn{9}{c}{Channels}\\
    $J^{PC}$    &1 &2  &3  &4  &5  &6  &7  &8  &9\\
  \midrule[1pt]
  $0^{++}$  &$D_1\bar{D}_1(|{}^1S_0\rangle)$
            &$D_1\bar{D}_1(|{}^5D_0\rangle)$
            &$D_2^*\bar{D}_2^*(|{}^1S_0\rangle)$
            &$D_2^*\bar{D}_2^*(|{}^5D_0\rangle)$    \\
  %%%%%%%%%%%%%%%
  $1^{+-}$  &$D_1\bar{D}_1(|{}^3S_1\rangle)$
            &$D_1\bar{D}_1(|{}^3D_1\rangle)$
         &$D_1\bar{D}_2^*(|{}^3S_1\rangle)$
         &$D_1\bar{D}_2^*(|{}^3D_1\rangle)$
         &$D_2^*\bar{D}_2^*(|{}^3S_1\rangle)$
         &$D_2^*\bar{D}_2^*(|{}^3D_1\rangle)$   \\
  %%%%%%%%%%%%%%%
  $2^{++}$  &$D_1\bar{D}_1(|{}^5S_2\rangle)$~\,
            &$D_1\bar{D}_1(|{}^1D_2\rangle)$~\,
            &$D_1\bar{D}_1(|{}^5D_2\rangle)$~\,
         &$D_1\bar{D}_2^*(|{}^5S_2\rangle)$~\,
         &$D_1\bar{D}_2^*(|{}^3D_2\rangle)$~\,
         &$D_1\bar{D}_2^*(|{}^5D_2\rangle)$~\,
         &$D_2^*\bar{D}_2^*(|{}^5S_2\rangle)$~\,
         &$D_2^*\bar{D}_2^*(|{}^1D_2\rangle)$~\,
         &$D_2^*\bar{D}_2^*(|{}^5D_2\rangle)$ \\
  %%%%%%%%%%%%%%%
  $3^{+-}$  &$D_1\bar{D}_2^*(|{}^7S_3\rangle)$
            &$D_1\bar{D}_2^*(|{}^3D_3\rangle)$
            &$D_1\bar{D}_2^*(|{}^7D_3\rangle)$
         &$D_2^*\bar{D}_2^*(|{}^7S_3\rangle)$
         &$D_2^*\bar{D}_2^*(|{}^3D_3\rangle)$
         &$D_2^*\bar{D}_2^*(|{}^7D_3\rangle)$  \\
\bottomrule[1pt] \bottomrule[1pt]
\end{tabular}
\end{table*}

\renewcommand{\arraystretch}{1.5}
\begin{table*}[!hbtp]
\caption{Probabilities of each channel for the selected pure
$D_1\bar{D}_2^*$ and $D_2^*\bar{D}_2^*$ systems. The notation
$\cdots$ means that no binding solution is found.}\label{poss}
\begin{tabular}{cllllllll}
\toprule[1pt]\toprule[1pt]
\multirow{2}*{$D_1\bar{D}_2^*$}  &\multicolumn{8}{c}{$I^G(J^{PC})$} \\
&\multicolumn{2}{c}{$0^-(1^{+-})$}
&\multicolumn{2}{c}{$0^+(2^{++})$}
&\multicolumn{2}{c}{$0^-(3^{+-})$}
&\multicolumn{2}{c}{$1^+(3^{+-})$}
\\\midrule[1pt]
$\Lambda$ (GeV)   &&1.42   &&2.40    &&3.13   &&4.81\\
S-wave contribution (\%)      &$|{}^3S_1\rangle$ &99.68 \,\,\,\,\,\,
   &$|{}^5S_2\rangle$  &97.38\,\,\,\,\,\,    &$|{}^7S_3\rangle$ &92.52 \,\,\,\,\,\,    &$|{}^7S_3\rangle$  &98.30\\
D-wave contribution (\%)
  &$|{}^3D_1\rangle$   &0.05
  &$|{}^3D_2\rangle$ &$\sim0$  &$|{}^3D_3\rangle$ &0.16
  &$|{}^3D_3\rangle$   &0.51\\
  &$|{}^5D_1\rangle$ &$\sim0$  &$|{}^5D_2\rangle$ &2.62
  &$|{}^5D_3\rangle$ &$\sim0$  &$|{}^5D_3\rangle$  &$\sim0$ \\
  &$|{}^7D_1\rangle$ &0.27     &$|{}^7D_2\rangle$  &$\sim0$  &$|{}^7D_3\rangle$ &7.33     &$|{}^7D_3\rangle$  &1.54\\
                \midrule[1pt]
\multirow{2}*{$D_2^*\bar{D}_2^*$}  &\multicolumn{8}{c}{$I^G(J^{PC})$}      \\
&\multicolumn{2}{c}{$0^-(1^{+-})$}
&\multicolumn{2}{c}{$0^+(2^{++})$}
&\multicolumn{2}{c}{$0^-(3^{+-})$}
&\multicolumn{2}{c}{$1^+(3^{+-})$}
\\\midrule[1pt]
$\Lambda$ (GeV)     &&1.32     &&1.84     &&2.90    &&\ldots\\
S-wave contribution (\%)
  &$|{}^3S_1\rangle$ &98.53    &$|{}^5S_2\rangle$ &97.74    &$|{}^7S_3\rangle$ &94.25    &$|{}^7S_3\rangle$ &\ldots\\
D-wave contribution (\%)
  &$|{}^3D_1\rangle$ &0.90     &$|{}^1D_2\rangle$ &0.04     &$|{}^3D_3\rangle$  &0.54     &$|{}^3D_3\rangle$  &\ldots\\
  &$|{}^7D_1\rangle$  &0.58    &$|{}^5D_2\rangle$ &1.75     &$|{}^7D_3\rangle$   &5.21    &$|{}^7D_3\rangle$  &\ldots\\
  &                   &        &$|{}^9D_2\rangle$ &0.48     &                    &&\\
\bottomrule[1pt] \bottomrule[1pt]
\end{tabular}
\end{table*}

To get the final potentials, one repeats the procedure to obtain Eq.
(\ref{D1D1pot}) with the relevant transition amplitudes. For the
$J^{PC}=0^{++}$ case, we have
\begin{eqnarray}
&&\mathcal{V}(I,{J=0},r)=\nonumber\\
&& \scriptsize{\left(\begin{array}{cc}\langle
X_{D_1\bar{D}_1}|{V}\left[D_1\bar{D}_1\to
D_1\bar{D}_1\right](r)|X_{D_1\bar{D}_1}\rangle
       &\langle X_{D_1\bar{D}_1}|{V}\left[D_1\bar{D}_1\to D_2^*\bar{D}_2^*\right](r)|X_{D_2^*\bar{D}_2^*}\rangle\\
       \langle X_{D_2^*\bar{D}_2^*}|{V}\left[D_2^*\bar{D}_2^*\to D_1\bar{D}_1\right](r)|X_{D_1\bar{D}_1}\rangle
       &\langle X_{D_2^*\bar{D}_2^*}|{V}\left[D_2^*\bar{D}_2^*\to D_2^*\bar{D}_2^*\right](r)|X_{D_2^*\bar{D}_2^*}\rangle
       \end{array}\right)}\nonumber\\
        &&=\mathcal{G}(I)
         \left(\begin{array}{cc}\mathcal{V}_{1}        &\mathcal{V}_{4}\\
                                \mathcal{V}_{4}        &\mathcal{V}_{10}
         \end{array}\right),\label{totalpential4}
\end{eqnarray}
where $\mathcal{V}_4$ is a 2 by 2 matrix. The base of the whole
$4\times 4$ matrix is the column matrix composed of the four scalar
states in Table \ref{channel}.

Similarly, the final potentials for the cases $J^{PC}=1^{+-}$,
$2^{++}$, and $3^{+-}$ are
\begin{eqnarray}
&&\mathcal{V}(I,J=1,r)\nonumber\\
 &&= \mathcal{G}(I)\scriptsize{\left(\begin{array}{ccc}
                 \mathcal{V}_{1}       &\frac{1}{\sqrt{2}}\mathcal{V}_{2}+\frac{c}{\sqrt{2}}\mathcal{V}_{3}     &\mathcal{V}_{4}  \\
                 \frac{1}{\sqrt{2}}\mathcal{V}_{2}+\frac{c}{\sqrt{2}}\mathcal{V}_{3}       &\mathcal{V}_{5}+c\mathcal{V}_{6}     &\frac{1}{\sqrt{2}}\mathcal{V}_{7}+\frac{c}{\sqrt{2}}\mathcal{V}_{9}   \\
                 \mathcal{V}_{4}       &\frac{1}{\sqrt{2}}\mathcal{V}_{7}+\frac{c}{\sqrt{2}}\mathcal{V}_{9}     &\mathcal{V}_{10}
         \end{array}\right)},\label{totalpential5}
\end{eqnarray}
\begin{eqnarray}
&&\mathcal{V}(I,J=2,r)\nonumber\\
&&= \mathcal{G}(I)\scriptsize{\left(\begin{array}{ccc}
                 \mathcal{V}_{1}       &\frac{1}{\sqrt{2}}\mathcal{V}_{2}-\frac{c}{\sqrt{2}}\mathcal{V}_{3}     &\mathcal{V}_{4}  \\
                 \frac{1}{\sqrt{2}}\mathcal{V}_{2}-\frac{c}{\sqrt{2}}\mathcal{V}_{3}       &\mathcal{V}_{5}-c\mathcal{V}_{6}     &\frac{1}{\sqrt{2}}\mathcal{V}_{7}-\frac{c}{\sqrt{2}}\mathcal{V}_{9}   \\
                 \mathcal{V}_{4}       &\frac{1}{\sqrt{2}}\mathcal{V}_{7}-\frac{c}{\sqrt{2}}\mathcal{V}_{9}     &\mathcal{V}_{10}
         \end{array}\right)},\label{totalpential6}
\end{eqnarray}
and
\begin{eqnarray}
&&\mathcal{V}(I,J=3,r)\nonumber\\
 &&= \mathcal{G}(I)\left(\begin{array}{cc}
                 \mathcal{V}_{5}+c\mathcal{V}_{6}    &\frac{1}{\sqrt{2}}\mathcal{V}_{7}+\frac{c}{\sqrt{2}}\mathcal{V}_{9}\\
                 \frac{1}{\sqrt{2}}\mathcal{V}_{7}+\frac{c}{\sqrt{2}}\mathcal{V}_{9}   &\mathcal{V}_{10}
         \end{array}\right),\label{totalpential7}
\end{eqnarray}
respectively. The bases of these potential matrices (also the order
of elements) are completely determined by the channels in Table
\ref{channel}. Since we ignore small contributions from several
channels, the dimension of $\mathcal{V}_{5,6,10}$ here is smaller
than that in Eqs. (\ref{totalpential2}) and (\ref{totalpential3}).
Note we do not give $\mathcal{V}_8$ and $\mathcal{V}_{11}$
explicitly as in Eq. (\ref{totalpential2}), since they are equal to
$\mathcal{V}_5$ and $\mathcal{V}_6$, respectively. One may find the
expressions of these eleven potentials $\mathcal{V}_k$ ($k=1,2,
\cdots, 11$) in the Appendix \ref{app01}.

Following the same procedure to solve the bound state problem, one
gets the possible binding energies and RMS radius. In Table
\ref{couple}, we present the numerical results for the coupled
$T\bar{T}$ systems. Probabilities ($p_i$ (\%)) for each channel are
also given. There are six bound state solutions with the quantum
numbers $I^G(J^{PC})=0^+(0^{++})$, $0^-(1^{+-})$, $0^+(2^{++})$,
$1^-(2^{++})$, $0^-(3^{+-})$ and $1^+(3^{+-})$. In Fig. \ref{cp}, we
illustrate the cutoff dependence for the binding energy $E$ and the
RMS radius $r_{RMS}$.

\begin{figure}[!htbp]
  \centering
  \includegraphics[width=3.4in]{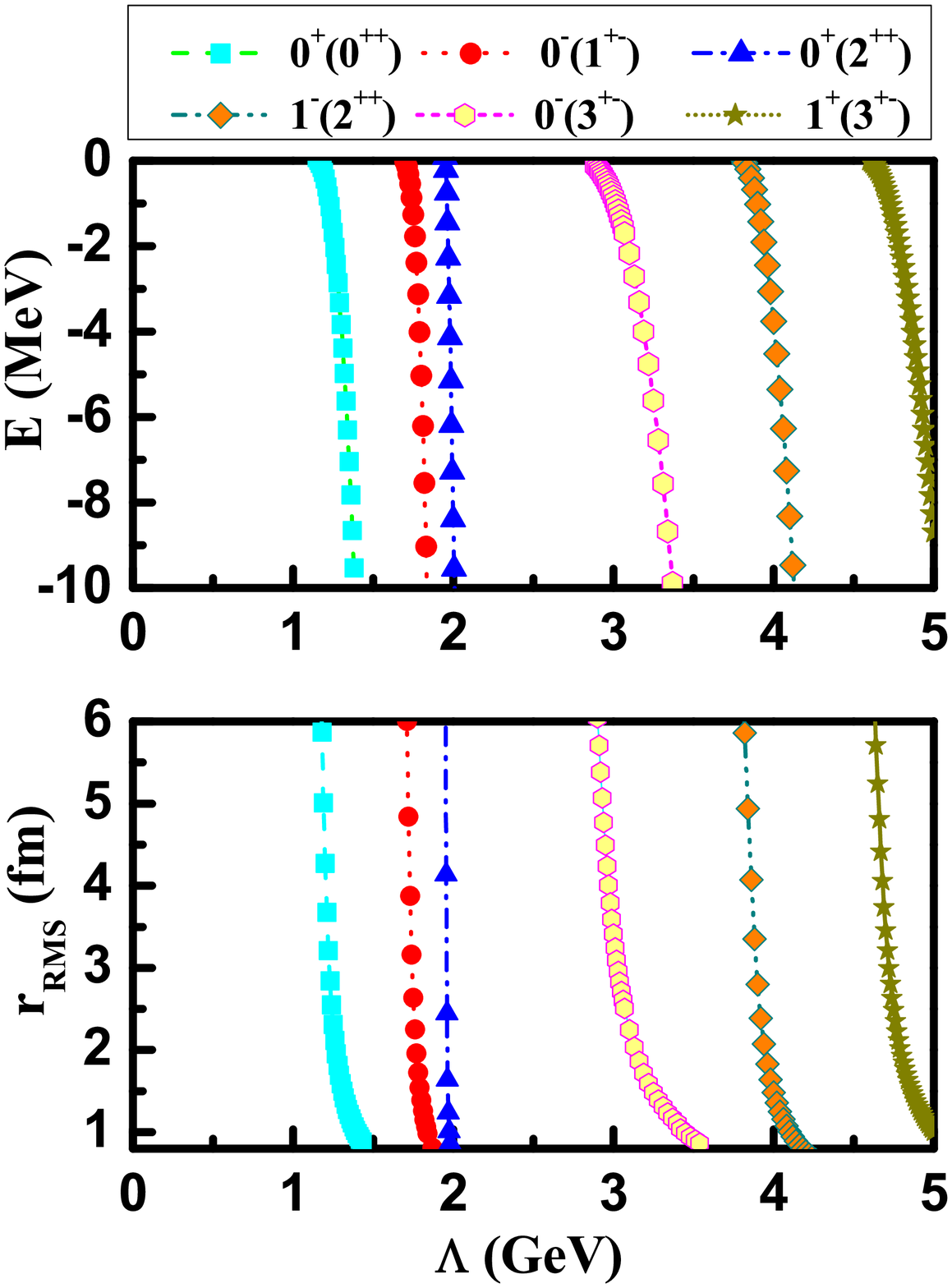}\\
\caption{(color online). The $\Lambda$ dependence of the bound state
solutions (the binding energy $E$ and the root-mean-square radius
$r_{RMS}$) for the $T\bar{T}$ systems with the coupled channel
effect.}\label{cp}
\end{figure}

\renewcommand{\arraystretch}{1.5}
\begin{table*}[!hbtp]
\caption{The binding solutions (binding energy $E$ and
root-mean-square radius $r_{RMS}$) after the coupled channel effects
are considered. $p_{i}(\%)$ denotes the probability for the $i$-th
channel. $E$, $r_{RMS}$, and $\Lambda$ are in units of MeV, fm, and
GeV, respectively. The notation $\cdots$ means that no binding
solution is found.}\label{couple}
\begin{tabular}{c|ccc|ccccccccc}
\toprule[1pt]\toprule[1pt] $I^G(J^{PC})$      &$\Lambda$    &$E$
&$r_{RMS}$
      &$p_1 (\%)$   &$p_2 (\%)$     &$p_3 (\%)$      &$p_4 (\%)$     &$p_5 (\%)$                &$p_6 (\%)$   &$p_7 (\%)$     &$p_8 (\%)$      &$p_9 (\%)$\\\midrule[1pt]
   &&&
  &$D_1\bar{D}_1|{}^1S_0\rangle$         &$D_1\bar{D}_1|{}^5D_0\rangle$
  &$D_2^*\bar{D}_2^*|{}^1S_0\rangle$     &$D_2^*\bar{D}_2^*|{}^5D_0\rangle$\\
$0^+(0^{++})$       &1.30               &-3.84         &1.59
      &97.91        &0.82           &1.26            &0.01     &\\
                   &1.40               &-11.46        &0.98
      &95.46        &0.90     &3.60     &0.03&\\
$1^-(0^{++})$   &\ldots&\ldots&\ldots
      &\ldots&\ldots&\ldots&\ldots&\\\hline
     &&&
      &$D_1\bar{D}_1|{}^3S_1\rangle$    &$D_1\bar{D}_1|{}^3D_1\rangle$
      &$D_1\bar{D}_2^*|{}^3S_1\rangle$  &$D_1\bar{D}_2^*|{}^3D_1\rangle$
      &$D_2^*\bar{D}_2^*|{}^3S_1\rangle$   &$D_2^*\bar{D}_2^*|{}^3D_1\rangle$\\
$0^-(1^{+-})$       &1.78      &-3.13      &1.73      &82.66
&1.44     &$\sim 0$
      &$\sim 0$       &15.70     &0.20    &\\
                   &1.84      &-10.69     &0.98      &60.04     &1.26   &$\sim 0$
      &$\sim 0$       &38.16    &0.54     &\\
$1^+(1^{+-})$   &\ldots&\ldots&\ldots
      &\ldots&\ldots&\ldots&\ldots&\ldots&\ldots&\\\hline
      &&&
      &$D_1\bar{D}_1|{}^5S_2\rangle$
      &$D_1\bar{D}_1|{}^1D_2\rangle$
      &$D_1\bar{D}_1|{}^5D_2\rangle$
         &$D_1\bar{D}_2^*|{}^5S_2\rangle$
         &$D_1\bar{D}_2^*|{}^3D_2\rangle$
         &$D_1\bar{D}_2^*|{}^5D_2\rangle$
              &$D_2^*\bar{D}_2^*|{}^5S_2\rangle$
              &$D_2^*\bar{D}_2^*|{}^1D_2\rangle$
              &$D_2^*\bar{D}_2^*|{}^5D_2\rangle$ \\
$0^+(2^{++})$      &1.97      &-2.27      &1.24      &30.46
&0.08    &1.00
      &34.53       &$\sim 0 $    &0.13    &33.19     &0.27     &0.34\\
      &1.98        &-4.13     &0.86       &22.28     &0.07     &0.87     &38.04
      &$\sim 0 $    &0.14     &37.91     &0.31    &0.39\\
$1^-(2^{++})$      &3.96      &-2.45      &1.83      &94.91
&0.20    &1.12
      &3.11      &0.10      &$\sim 0$    &0.49    &$\sim 0$&$\sim 0$\\
      &4.12     &-9.46    &0.95    &90.78    &0.31    &1.70    &5.92    &0.17
      &$\sim 0$   &1.00&$\sim 0$&$\sim 0$\\\hline
      &&&
      &$D_1\bar{D}_2^*|{}^7S_3\rangle$
      &$D_1\bar{D}_2^*|{}^3D_3\rangle$
      &$D_1\bar{D}_2^*|{}^7D_3\rangle$
          &$D_2^*\bar{D}_2^*|{}^7S_3\rangle$
          &$D_2^*\bar{D}_2^*|{}^3D_3\rangle$
          &$D_2^*\bar{D}_2^*|{}^7D_3\rangle$\\
$0^-(3^{+-})$     &3.16      &-3.31      &1.86      &91.90    &0.16
&7.94
      &$\sim 0 $&$\sim 0 $&$\sim 0 $&\\
      &3.37        &-9.89     &1.17     &87.97     &0.21     &11.82
      &$\sim 0 $&$\sim 0 $&$\sim 0 $&\\
$1^+(3^{+-})$     &4.79    &-2.08    &2.02    &98.43    &0.14
&1.43
      &$\sim 0 $&$\sim 0 $&$\sim 0 $&\\
      &4.98     &-7.85     &1.07     &97.48     &0.23     &2.33
      &$\sim 0 $&$\sim 0 $&$\sim 0 $&\\
\bottomrule[1pt]\bottomrule[1pt]
\end{tabular}
\end{table*}

The coupled system with $J=0$ includes four channels
$D_1\bar{D}_1(|{}^1S_0\rangle)$, $D_1\bar{D}_1(|{}^5D_0\rangle)$,
$D_2^*\bar{D}_2^*(|{}^1S_0\rangle)$ and
$D_2^*\bar{D}_2^*(|{}^5D_0\rangle)$. It is a good molecular
candidate since the cutoff parameter $\Lambda$ is around 1 GeV and
consistent with the experience from the deuteron. The mass of the
candidate is expected to be below the $D_1\bar{D}_1$ threshold
(around 4840 MeV). Moreover, it is a typical $S$-wave bound state as
the channel $D_1\bar{D}_1(|{}^1S_0\rangle)$ dominates exclusively
and the $D_2^*\bar{D}_2^*$ channels have small contributions around
several percents.

Six channels are considered for the state with $0^-(1^{+-})$:
$D_1\bar{D}_1(|{}^3S_1\rangle)$, $D_1\bar{D}_1(|{}^3D_1\rangle)$,
$D_1\bar{D}_2^*(|{}^3S_1\rangle)$,
$D_1\bar{D}_2^*(|{}^3D_1\rangle)$,
$D_2^*\bar{D}_2^*(|{}^3S_1\rangle)$ and
$D_2^*\bar{D}_2^*(|{}^3D_1\rangle)$. There are small components of
$D_1\bar{D}_2^*$ since the probabilities of
$D_1\bar{D}_2^*(|{}^3S_1\rangle)$ ($p_3$) and
$D_1\bar{D}_2^*(|{}^3D_1\rangle)$ ($p_4$) are both around zero even
with a larger binding energy about 10 MeV. For the remaining four
channels, the contributions from S-wave
$D_1\bar{D}_1(|{}^3S_1\rangle)$ and
$D_2^*\bar{D}_2^*(|{}^3S_1\rangle)$ play a leading role in this
state. With the binding energy becoming larger and larger, the
$D_2^*\bar{D}_2^*$ channel will play a more significant and
increasing role. However, the chance for the $0^-(1^{+-})$ state to
be detected by experiments may be lower than that for the
$0^+(0^{++})$ as the cutoff
$\Lambda\left[0^-(1^{+-})\right]>\Lambda\left[0^+(0^{++})\right]$
when the binding energies are fixed at the same level.

For the state with $0^+(2^{++})$, the coupled channel effects due to
$D_1\bar{D}_2^*$ and $D_2^*\bar{D}_2^*$ are important since the
probabilities for $D_1\bar{D}_1$, $D_1\bar{D}_2^*$ and
$D_2^*\bar{D}_2^*$ are relatively close. The dominant channels are
all S-wave $|{}^5S_2\rangle$ with probability around $99\%$. It can
also be a good molecular candidate since the cutoff is around 2 GeV.

The above three solutions are all isoscalar states and the masses
are all around the $D_1\bar{D}_1$ threshold. The general feature
does not change compared to the previous case but the coupled
channel effects lower the binding energies. For the remaining cases,
the solutions with $1^-(2^{++})$, $0^-(3^{+-})$ and $1^+(3^{+-})$
correspond to the typical $D_1\bar{D}_1$, $D_1\bar{D}_2^*$ and
$D_1\bar{D}_2^*$ bound state, since their dominant channels are
$D_1\bar{D}_1(|{}^5S_2\rangle)$, $D_1\bar{D}_2^*(|{}^7S_3\rangle)$
and $D_1\bar{D}_2^*(|{}^7S_3\rangle)$, respectively. However, the
corresponding cutoff is far from the usual value around 1 GeV.

\section{Summary}\label{sec4}

In this work, we have studied the interactions of the
$T\bar{T}$-type molecular systems within the framework of the
one-pion-exchange model. We have found the bound state solutions for
some $T\bar{T}$-type molecular systems. There exist possible
$T\bar{T}$-type molecular states, which are summarized in Table
\ref{state} for readers' convenience.

\renewcommand{\arraystretch}{1.5}
\begin{table}[!hbtp]
\caption{Summary of the $T\bar{T}$ systems. }\label{state}
\begin{tabular}{ccc|ccc|cc}
\toprule[1pt]\toprule[1pt]
\multicolumn{3}{c}{Pure system}&\multicolumn{3}{c}{Pure system}  &\multicolumn{2}{c}{Coupled system}\\
\midrule[1pt]
      &$I^G(J^{PC})$   &Remark    &     &$I^G(J^{PC})$   &Remark  &$I^G(J^{PC})$   &Remark
\\\midrule[1pt]
$D_1\bar{D}_1$
      &$0^+(0^{++})$    &$\star\star\star\star$
&$D_2^*\bar{D}_2^*$
      &$0^+(0^{++})$    &$\star\star\star\star$
      &$0^+(0^{++})$    &$\star\star\star\star$\\

      &$0^-(1^{+-})$    &$\star\star\star$
      &&$0^-(1^{+-})$   &$\star\star\star\star$
      &$0^-(1^{+-})$    &$\star\star\star$\\

      &$0^+(2^{++})$    &$\star$
      &&$0^+(2^{++})$   &$\star\star\star$
      &$0^+(2^{++})$    &$\star\star\star$\\

      &$1^-(2^{++})$    &$\star$
      &&$0^-(3^{+-})$   &$\star\star$
      &$1^-(2^{++})$    &$\star$\\

      &&&&$0^+(4^{++})$   &$\star$
      &$0^-(3^{+-})$   &$\star\star$ \\

      &&&&$1^-(4^{++})$   &$\star$
      &$1^+(3^{+-})$   &$\star$ \\\midrule[1pt]

$D_1\bar{D}_2^*$
      &$0^-(1)^{+-}$   &$\star\star\star\star$
&$D_1\bar{D}_2^*$
      &$0^+(1)^{++}$   &$\star\star\star\star$\\
      &$0^-(2)^{+-}$   &$\star\star$
      &&$0^+(2)^{++}$  &$\star\star\star$\\
      &$0^-(3)^{+-}$   &$\star\star$
      &&$0^+(3)^{++}$   &$\star$\\
      &$1^+(3)^{+-}$   &$\star$
      &&$1^-(3)^{++}$   &$\star$\\
\bottomrule[1pt]\bottomrule[1pt]
\end{tabular}
\end{table}

In Table \ref{state}, we adopt the same criterion as that in Ref.
\cite{Sun:2012sy} and show the existence possibility of these
bound-state solutions with asterisks. A state with more $\star$
implies higher possibility to find this molecular state. Here, the
four-star, three-star, two-star and one-star are applied to mark
these states, which have bound state solutions with cutoff
$\Lambda<1.5$ GeV, $1.5<\Lambda<2.5$ GeV, $2.5<\Lambda<3.5$ GeV and
$3.5<\Lambda<5$ GeV, respectively. Thus, we suggest experiments to
focus on these four-star states, which include the $D_1\bar{D}_1$
state with $I^G(J^{PC})=0^+(0^{++})$, the $D_1\bar{D}_2^*$ states
with $0^{\pm}(1)^{+\pm}$, the $D_2^*\bar{D}_2^*$ state with
$0^+(0^{++})$ and the $0^+(0^{++})$
$D_1\bar{D}_1-D_1\bar{D}_2^*-D_2^*\bar{D}_2^*$ coupled state. In
other words, one or several resonant structures around the
$D_1\bar{D}_1$ threshold are highly probable.

In the present model we adopt a phenomenological form factor at each
interaction vertex with the adjustable cutoff parameter $\Lambda$.
The numerical results are sensitive to its value. Because of lack of
available experimental data, we restrict its reasonable range from
the experience with the deuteron. The numerical results are
preliminary. In principle, the results should be stable with the
variation of the cutoff. Probably the consideration of the other
meson exchange forces may reduce the sensitivity once the coupling
constants can be appropriately determined. Hopefully one can improve
the model and make more reliable predictions in the future.

Searching for the exotic multiquark states continues to be a very
interesting issue of hadron physics. The present predictions of the
hidden-charm molecular states, which are composed of anti-charmed
and charmed meson in the $T$ doublet, provide useful information for
future experimental exploration of them. The present work is only
the starting point for the study of the $T\bar{T}$-type molecular
systems, which need to be investigated further by other approaches.

Recent experimental observation of the two hidden-charm pentaquarks
$P_c(4380)$ and $P_c(4450)$ \cite{Aaij:2015tga} enhances our
confidence in the existence of the multiquark states. With
experimental progresses, especially from LHCb and forthcoming
BelleII, we expect that more candidates of the multiquark states
will be announced. This field is full of challenges and
opportunities for both theorists and experimentalists.

\appendix

\section{Relevant subpotentials}\label{app01}
We list the subpotentials
$\hat{\mathcal{V}}_{1}-\hat{\mathcal{V}}_{11}$ used in the paper.
For convenience, we define
\begin{eqnarray}
\begin{array}{ll}
\Lambda_1 =\sqrt{ \Lambda^2-(m_{D_2^*}-m_{D_1})^2},
&m_{\pi1} =\sqrt{ m_{\pi}^2-(m_{D_2^*}-m_{D_1})^2},\\
\Lambda_2 =\sqrt{
\Lambda^2-\left(\frac{m_{D_2^*}^2-m_{D_1}^2}{4m_{D_1}}\right)^2},
&m_{\pi2} =\sqrt{ m_{\pi}^2-\left(\frac{m_{D_2^*}^2-m_{D_1}^2}{4m_{D_1}}\right)^2},\\
\Lambda_3 =\sqrt{
\Lambda^2-\left(\frac{m_{D_2^*}^2-m_{D_1}^2}{4m_{D_2^*}}\right)^2},
&m_{\pi3} =\sqrt{
m_{\pi}^2-\left(\frac{m_{D_2^*}^2-m_{D_1}^2}{4m_{D_2^*}}\right)^2},
\end{array}\nonumber\\
\end{eqnarray}
and
\begin{eqnarray}
Y(\Lambda,m,{r}) &=&
       \frac{1}{4\pi r}(e^{-mr}-e^{-\Lambda r})-\frac{\Lambda^2-m^2}{8\pi \Lambda}e^{-\Lambda r},\nonumber\\
S(\hat{r},\vec{a},\vec{b}) &=&
3(\hat{r}\cdot\vec{a})(\hat{r}\cdot\vec{b})-\vec{a}\cdot\vec{b},
\end{eqnarray}
with $\hat{r}=\vec{r}/|\vec{r}|$.

The first potential is for the $D_1\bar{D}_1$ system,
\begin{eqnarray}\label{poten1}
\hat{\mathcal{V}}_{1} &\equiv& \mathcal{V}\left[{D_1(1)\bar{D}_1(2)\rightarrow D_1(3)\bar{D}_1}(4)\right]({r})\nonumber\\
&=&
\frac{25}{108}\frac{k^2}{f_{\pi}^2}\Bigg\{\mathcal{E}_1\nabla^2+\mathcal{S}_1
         r\frac{d}{d r}\frac{1}{r}\frac{d}{d r}}{\Bigg\}Y(\Lambda,m_{\pi},r),
\end{eqnarray}
where the spin-spin operator is defined as
$\mathcal{E}_1=(\vec{\epsilon}_1\times\vec{\epsilon}_3^{\dag})\cdot
(\vec{\epsilon}_2\times\vec{\epsilon}_4^{\dag})$ and the tensor
operator is $\mathcal{S}_1=
S(\hat{r},\vec{\epsilon}_1\times\vec{\epsilon}_3^{\dag},\vec{\epsilon}_2\times\vec{\epsilon}_4^{\dag})$.
$\mathcal{V}$, $\mathcal{E}$, and $\mathcal{S}$ have the same
subscript while the subscripts 1, 2, 3, and 4 of the polarization
vectors refer to the incoming meson (1), incoming meson (2),
outgoing meson (3), and outgoing meson (4) in the t-channel
scattering process, respectively.

The other potentials are
\begin{eqnarray}
\hat{\mathcal{V}}_{2} &=& \mathcal{V}\left[{D_1(1)\bar{D}_1(2)\rightarrow D_1(3)\bar{D}_2^*}(4)\right]({r})\nonumber\\
                &=& -i\frac{5}{18\sqrt{6}}\frac{k^2}{f_{\pi}^2}
                \Bigg\{\mathcal{E}_2\nabla^2+\mathcal{S}_2 r\frac{d}{d r}\frac{1}{r}\frac{d}{d r}\Bigg\}
                Y(\Lambda_2,m_{\pi2},r),\label{poten2}\nonumber\\
\end{eqnarray}
with $\mathcal{E}_2=\sum_{c,d}C_{1,c;1,d}^{2,c+d}
(\vec{\epsilon}_2\cdot\vec{\epsilon}_{4c}^{\dag})(\vec{\epsilon}_1\times\vec{\epsilon}_3^{\dag})\cdot\vec{\epsilon}_{4d}^{\dag}$
and $\mathcal{S}_2=\sum_{c,d}C_{1,c;1,d}^{2,c+d}
(\vec{\epsilon}_2\cdot\vec{\epsilon}_{4c}^{\dag})S(\hat{r},\vec{\epsilon}_1\times\vec{\epsilon}_3^{\dag},\vec{\epsilon}_{4d}^{\dag})$,

\begin{eqnarray}
\hat{\mathcal{V}}_{3} &=& \mathcal{V}\left[{D_1(1)\bar{D}_1(2)\rightarrow D_2^*(3)\bar{D}_1(4)}\right]({r})\nonumber\\
                &=& i\frac{5}{18\sqrt{6}}\frac{k^2}{f_{\pi}^2}\Bigg\{\mathcal{E}_3\nabla^2+\mathcal{S}_3r\frac{d}{d r}\frac{1}{r}\frac{d}{d r}\Bigg\}Y(\Lambda_2,m_{\pi2},r),\label{poten3}\nonumber\\
\end{eqnarray}
with $\mathcal{E}_3=\sum_{c,d}C_{1,c;1,d}^{2,c+d}
(\vec{\epsilon}_1\cdot\vec{\epsilon}_{3c}^{\dag})(\vec{\epsilon}_2\times\vec{\epsilon}_4^{\dag})\cdot\vec{\epsilon}_{3d}^{\dag}$
and $\mathcal{S}_3=\sum_{c,d}C_{1,c;1,d}^{2,c+d}
(\vec{\epsilon}_1\cdot\vec{\epsilon}_{3c}^{\dag})S(\hat{r},\vec{\epsilon}_2\times\vec{\epsilon}_4^{\dag},\vec{\epsilon}_{3d}^{\dag})$,

\begin{eqnarray}
\hat{\mathcal{V}}_{4} &=& \mathcal{V}\left[{D_1(1)\bar{D}_1(2)\rightarrow D_2^*(3)\bar{D}_2^*(4)}\right]({r})\nonumber\\
&=&
\frac{1}{18}\frac{k^2}{f_{\pi}^2}\Bigg\{\mathcal{E}_4\nabla^2+\mathcal{S}_4r\frac{d}{d
r}\frac{1}{r}\frac{d}{d r}\Bigg\}
Y(\Lambda,m_{\pi},r),\label{poten4}
\end{eqnarray}
with $\mathcal{E}_4=\sum_{c,d,i,j}
 C_{1,c;1,d}^{2,c+d}C_{1,m;1,n}^{2,m+n} (\vec{\epsilon}_1\cdot\vec{\epsilon}_{3c}^{\dag})
(\vec{\epsilon}_2\cdot\vec{\epsilon}_{4m}^{\dag})
(\vec{\epsilon}_{3d}^{\dag}\cdot\vec{\epsilon}_{4n}^{\dag})
\cdot\vec{\epsilon}_{3d}^{\dag}$ and
$\mathcal{S}_4=\sum_{c,d,i,j}C_{1,c;1,d}^{2,c+d}C_{1,m;1,n}^{2,m+n}
(\vec{\epsilon}_1\cdot\vec{\epsilon}_{3c}^{\dag})
(\vec{\epsilon}_2\cdot\vec{\epsilon}_{4m}^{\dag})
S(\hat{r},\vec{\epsilon}_{3d}^{\dag},\vec{\epsilon}_{4n}^{\dag})$,

\begin{eqnarray}
\hat{\mathcal{V}}_{5} &=& \mathcal{V}\left[{D_1(1)\bar{D}_2^*(2)\rightarrow D_1(1)\bar{D}_2^*(2)}\right]({r})\nonumber\\
                &=& \frac{5}{18}\frac{k^2}{f_{\pi}^2}
                \Bigg\{\mathcal{E}_5\nabla^2+\mathcal{S}_5
                r\frac{d}{d r}\frac{1}{r}\frac{d}{d r}\Bigg\}Y(\Lambda,m_{\pi},r),\label{poten5}
\end{eqnarray}
with
$\mathcal{E}_5=\sum_{c,d,m,n}C_{1,c;1,d}^{2,c+d}C_{1,m;1,n}^{2,m+n}
(\vec{\epsilon}_{2c}\cdot\vec{\epsilon}_{4m}^{\dag})
[(\vec{\epsilon}_1\times\vec{\epsilon}_3^{\dag})\cdot
(\vec{\epsilon}_{2d}\times\vec{\epsilon}_{4n}^{\dag})]$ and
$\mathcal{S}_5=\sum_{c,d,m,n}C_{1,c;1,d}^{2,c+d}C_{1,m;1,n}^{2,m+n}
(\vec{\epsilon}_{2c}\cdot\vec{\epsilon}_{4m}^{\dag})
S(\hat{r},\vec{\epsilon}_1\times\vec{\epsilon}_3^{\dag},
\vec{\epsilon}_{2d}\times\vec{\epsilon}_{4n}^{\dag})$,

\begin{eqnarray}
\hat{\mathcal{V}}_{6} &=&
\mathcal{V}\left[{D_1(1)\bar{D}_2^*(2)\rightarrow D_2^*(3)\bar{D}_1(4)}\right]({r})\nonumber\\
                &=& \frac{1}{18}\frac{k^2}{f_{\pi}^2}
                \Bigg\{\mathcal{E}_6\nabla^2+\mathcal{S}_6r\frac{d}{d r}\frac{1}{r}\frac{d}{d r}\Bigg\} Y(\Lambda_1,m_{\pi1},r),\label{poten6}
\end{eqnarray}
with $\mathcal{E}_6
=\sum_{c,d,m,n}C_{1,c;1,d}^{2,c+d}C_{1,m;1,n}^{2,m+n}
(\vec{\epsilon}_1\cdot\vec{\epsilon}_{3c}^{\dag})
(\vec{\epsilon}_{2m}\cdot\vec{\epsilon}_{4}^{\dag})
(\vec{\epsilon}_{3d}^{\dag}\cdot\vec{\epsilon}_{2n})$ and
$\mathcal{S}_6=\sum_{c,d,m,n}C_{1,c;1,d}^{2,c+d}C_{1,m;1,n}^{2,m+n}
(\vec{\epsilon}_1\cdot\vec{\epsilon}_{3c}^{\dag})
(\vec{\epsilon}_{2m}\cdot\vec{\epsilon}_{4}^{\dag})
S(\hat{r},\vec{\epsilon}_{3d}^{\dag},\vec{\epsilon}_{2n})$,

\begin{eqnarray}
\hat{\mathcal{V}}_{7} &=& \mathcal{V}\left[{D_1(1)\bar{D}_2^*(2)\rightarrow D_2^*(3)\bar{D}_2^*(4)}\right]({r})\nonumber\\
                &=& \frac{i}{3\sqrt{6}}\frac{k^2}{f_{\pi}^2}
                \Bigg\{\mathcal{E}_7\nabla^2+\mathcal{S}_7 r\frac{d}{d r}\frac{1}{r}\frac{d}{d r}\Bigg\} Y(\Lambda_3,m_{\pi3},r),\label{poten7}\nonumber\\
\end{eqnarray}
with $\mathcal{E}_7
=\sum_{c,d,h,l,m,n}C_{1,c;1,d}^{2,c+d}C_{1,h;1,l}^{2,h+l}C_{1,m;1,n}^{2,m+n}
(\vec{\epsilon}_1\cdot\vec{\epsilon}_{3c}^{\dag})(\vec{\epsilon}_{2h}\cdot\vec{\epsilon}_{4m}^{\dag})
[\vec{\epsilon}_{3d}^{\dag}\cdot(\vec{\epsilon}_{2l}\times\vec{\epsilon}_{4n}^{\dag})]$
and
$\mathcal{S}_7=\sum_{c,d,m,n}C_{1,c;1,d}^{2,c+d}C_{1,m;1,n}^{2,m+n}
(\vec{\epsilon}_1\cdot\vec{\epsilon}_{3c}^{\dag})(\vec{\epsilon}_{2h}\cdot\vec{\epsilon}_{4m}^{\dag})
S(\hat{r},\vec{\epsilon}_{3d}^{\dag},\vec{\epsilon}_{2l}\times\vec{\epsilon}_{4n}^{\dag})$,

\begin{eqnarray}
\hat{\mathcal{V}}_{8} &=& \mathcal{V}\left[{D_2^*(1)\bar{D}_1(2)\rightarrow D_2^*(3)\bar{D}_1(4)}\right]({r})\nonumber\\
                &=& \frac{5}{18}\frac{k^2}{f_{\pi}^2}
                \Bigg\{\mathcal{E}_8\nabla^2+\mathcal{S}_8 r\frac{d}{d r}\frac{1}{r}\frac{d}{d r}\Bigg\} Y(\Lambda,m_{\pi},r),\label{poten8}\nonumber\\
\end{eqnarray}
with
$\mathcal{E}_8=\sum_{c,d,m,n}C_{1,c;1,d}^{2,c+d}C_{1,m;1,n}^{2,m+n}
(\vec{\epsilon}_{1c}\cdot\vec{\epsilon}_{3m}^{\dag})
[(\vec{\epsilon}_2\times\vec{\epsilon}_4^{\dag})\cdot
(\vec{\epsilon}_{1d}\times\vec{\epsilon}_{3n}^{\dag})]$ and
$\mathcal{S}_8=\sum_{c,d,m,n}C_{1,c;1,d}^{2,c+d}C_{1,m;1,n}^{2,m+n}
(\vec{\epsilon}_{1c}\cdot\vec{\epsilon}_{3m}^{\dag})
S(\hat{r},\vec{\epsilon}_2\times\vec{\epsilon}_4^{\dag},
\vec{\epsilon}_{1d}\times\vec{\epsilon}_{3n}^{\dag})$,

\begin{eqnarray}
\hat{\mathcal{V}}_{9} &=& \mathcal{V}\left[{D_2^*(1)\bar{D}_1(2)\rightarrow D_2^*(3)\bar{D}_2^*(4)}\right]({r})\nonumber\\
                &=& -i\frac{1}{3\sqrt{6}}\frac{k^2}{f_{\pi}^2}
                \Bigg\{\mathcal{E}_9\nabla^2+\mathcal{S}_9 r\frac{d}{d r}\frac{1}{r}\frac{d}{d r}\Bigg\} Y(\Lambda_3,m_{\pi3},r),\label{poten9}\nonumber\\
\end{eqnarray}
with $\mathcal{E}_9
=\sum_{c,d,h,l,m,n}C_{1,c;1,d}^{2,c+d}C_{1,h;1,l}^{2,h+l}C_{1,m;1,n}^{2,m+n}
(\vec{\epsilon}_2\cdot\vec{\epsilon}_{4c}^{\dag})(\vec{\epsilon}_{1h}\cdot\vec{\epsilon}_{3m}^{\dag})
[\vec{\epsilon}_{4d}^{\dag}\cdot(\vec{\epsilon}_{1l}\times\vec{\epsilon}_{3n}^{\dag})]$
and
$\mathcal{S}_9=\sum_{c,d,h,l,m,n}C_{1,c;1,d}^{2,c+d}C_{1,h;1,l}^{2,h+l}C_{1,m;1,n}^{2,m+n}
(\vec{\epsilon}_2\cdot\vec{\epsilon}_{4c}^{\dag})(\vec{\epsilon}_{1h}\cdot\vec{\epsilon}_{3m}^{\dag})
S(\hat{r},\vec{\epsilon}_{4d}^{\dag},
\vec{\epsilon}_{1l}\times\vec{\epsilon}_{3n}^{\dag})$,

\begin{eqnarray}
\hat{\mathcal{V}}_{10} &=& \mathcal{V}\left[{D_2^*(1)\bar{D}_2^*(2)\rightarrow D_2^*(3)\bar{D}_2^*(4)}\right]({r})\nonumber\\
                 &=& \frac{1}{3}\frac{k^2}{f_{\pi}^2}\Bigg\{\mathcal{E}_{10}\nabla^2+\mathcal{S}_{10} r\frac{\partial}{\partial r}\frac{1}{r}\frac{\partial}{\partial r}\Bigg\}
                 Y(\Lambda,m_{\pi},r),\label{poten10}\nonumber\\
\end{eqnarray}
with $\mathcal{E}_{10}
=\sum_{c,d,f,g,h,l,m,n}C_{1,c;1,d}^{2,c+d}C_{1,f;1,g}^{2,f+g}C_{1,h;1,l}^{2,h+l}C_{1,m;1,n}^{2,m+n}
(\vec{\epsilon}_{1c}\cdot\vec{\epsilon}_{3f}^{\dag})(\vec{\epsilon}_{2h}\cdot\vec{\epsilon}_{4m}^{\dag})
[(\vec{\epsilon}_{1d}\times\vec{\epsilon}_{3g}^{\dag})\cdot(\vec{\epsilon}_{2l}\times\vec{\epsilon}_{4n}^{\dag})]$
and $\mathcal{S}_{10}=\sum_{c,d,f,g,h,l,m,n}
C_{1,c;1,d}^{2,c+d}C_{1,f;1,g}^{2,f+g}C_{1,h;1,l}^{2,h+l}C_{1,m;1,n}^{2,m+n}
(\vec{\epsilon}_{1c}\cdot\vec{\epsilon}_{3f}^{\dag})(\vec{\epsilon}_{2h}\cdot\vec{\epsilon}_{4m}^{\dag})
S(\hat{r},\vec{\epsilon}_{1d}\times\vec{\epsilon}_{3g}^{\dag},\vec{\epsilon}_{2l}\times\vec{\epsilon}_{4n}^{\dag})$,

\begin{eqnarray}
\hat{\mathcal{V}}_{11} &=& \mathcal{V}\left[{D_2^*(1)\bar{D}_1(2)\rightarrow D_1(3)\bar{D}_2^*(4)}\right]({r})\nonumber\\
                &=& \frac{1}{18}\frac{k^2}{f_{\pi}^2}
                \Bigg\{\mathcal{E}_{11}\nabla^2+\mathcal{S}_{11}
                r\frac{d}{d r}\frac{1}{r}\frac{d}{d r}\Bigg\} Y(\Lambda_1,m_{\pi1},r),\label{poten11}\nonumber\\
\end{eqnarray}
with $\mathcal{E}_{11}
=\sum_{c,d,m,n}C_{1,c;1,d}^{2,c+d}C_{1,m;1,n}^{2,m+n}
(\vec{\epsilon}_2\cdot\vec{\epsilon}_{4c}^{\dag})(\vec{\epsilon}_{1m}\cdot\vec{\epsilon}_{3}^{\dag})
(\vec{\epsilon}_{4d}^{\dag}\cdot\vec{\epsilon}_{1n})$ and
$\mathcal{S}_{11}=
\sum_{c,d,m,n}C_{1,c;1,d}^{2,c+d}C_{1,m;1,n}^{2,m+n}
(\vec{\epsilon}_2\cdot\vec{\epsilon}_{4c}^{\dag})(\vec{\epsilon}_{1m}\cdot\vec{\epsilon}_{3}^{\dag})
S(\hat{r},\vec{\epsilon}_{4d}^{\dag},\vec{\epsilon}_{1n})$.

We present the matrix elements for the spin-spin operators
$\mathcal{E}_i$ and the generalized tensor operators $\mathcal{S}_i$
in Tables \ref{ss} and \ref{tensor}, respectively.

\renewcommand{\arraystretch}{2.0}
\begin{table*}[!htpb]
\caption{The matrix elements for the spin-spin operators
$\mathcal{E}_i$ in the effective potentials. Here, the subscript $i$
of $\mathcal{E}_i$ corresponds to that in the effective potentials
in Eqs. (\ref{poten1}-\ref{poten11}).}\label{ss}
\begin{tabular}{clllcclll}\toprule[1pt]\toprule[1pt]
$\langle\mathcal{E}_i[J]\rangle$
&\multicolumn{3}{c}{Matrices} &\quad\quad
&$\langle\mathcal{E}_i[J]\rangle$              &\multicolumn{3}{c}{Matrices}\\
\midrule[1pt]
$\langle\mathcal{E}_1[J]\rangle$        &\footnotesize{$\left(\begin{array}{cc}2  &0\\
                                    0  &-1\end{array}\right)_{J=0}$}
                           &\scriptsize{$\left(\begin{array}{cc}1   &0 \\
                                     0   &1\end{array}\right)_{J=1}$}
                     &{\scriptsize{$\left(\begin{array}{ccc}-1   &0    &0\\
                                      0    &2     &0\\
                                      0    &0     &-1\end{array}\right)_{J=2}$}}
                                      &\quad\quad
&$\langle\mathcal{E}_{2,3}[J]\rangle$    &\scriptsize{$\left(\begin{array}{cc}\sqrt{\frac{5}{6}}   &0\\
                                     0   &\sqrt{\frac{5}{6}}    \end{array}\right)_{J=1}$}
                           &\scriptsize{$\left(\begin{array}{ccc}
                                     \sqrt{\frac{3}{2}}    &0      &0\\
                                      0     &0    &0\\
                                      0  &0 &\sqrt{\frac{3}{2}}\end{array}\right)_{J=2}$}\\
$\langle\mathcal{E}_{4}[J]\rangle$      &\tiny{$\left(\begin{array}{cc}-\sqrt{\frac{5}{3}}  &0\\
                         0 &-\frac{\sqrt{\frac{7}{3}}}{2}\end{array}\right)_{J=0}$}
                           &\footnotesize{$\left(\begin{array}{cc}
                                     -\frac{\sqrt{5}}{2}   &0   \\
                                     0   &-\frac{\sqrt{5}}{2}\end{array}\right)_{J=1}$}
                           &{\tiny{$\left(\begin{array}{ccc}
                                    -\frac{\sqrt{\frac{7}{3}}}{2}   &0       &0\\
                                      0    &-\sqrt{\frac{5}{3}}       &0\\
                                      0    &0       &-\frac{\sqrt{\frac{7}{3}}}{2}
                                      \end{array}\right)_{J=2}$}}
                                      &\quad\quad
&$\langle\mathcal{E}_{5,8}[J]\rangle$
&\scriptsize{$\left(\begin{array}{cccc}
                                       \frac{3}{2} & 0 & 0 & 0 \\
                                        0 & \frac{3}{2} & 0 & 0 \\
                                        0 & 0 & \frac{1}{2} & 0 \\
                                        0 & 0 & 0 & -1
                                       \end{array}\right)_{J=1}$}
                           &\scriptsize{$\left(\begin{array}{cccc}
                                         \frac{1}{2} & 0 & 0 & 0 \\
                                         0  & \frac{3}{2} & 0 & 0 \\
                                         0 & 0  & \frac{1}{2} & 0 \\
                                         0 & 0 & 0  & -1
                                        \end{array}\right)_{J=2}$}
                           &{\scriptsize{$\left(\begin{array}{cccc}
                                         -1 & 0 & 0 & 0 \\
                                         0 & \frac{3}{2} & 0 & 0 \\
                                         0 & 0 & \frac{1}{2} & 0 \\
                                         0 & 0 & 0 & -1
                                        \end{array}\right)_{J=3}$}}\\

$\langle\mathcal{E}_{6,11}[J]\rangle$
&\scriptsize{$\left(\begin{array}{cccc}
                                          \frac{1}{6} & 0 & 0 & 0 \\
                                          0 & \frac{1}{6} & 0 & 0 \\
                                          0 & 0 & \frac{1}{2} & 0 \\
                                          0 & 0 & 0 & 1
                                         \end{array}\right)_{J=1}$}
                           &\scriptsize{$\left(\begin{array}{cccc}
                                           \frac{1}{2}  & 0 & 0 & 0 \\
                                           0  & \frac{1}{6} & 0 & 0 \\
                                           0  & 0 & \frac{1}{2} & 0 \\
                                           0  & 0 & 0 & 1
                                          \end{array}\right)_{J=2}$}
                           &{\scriptsize{$\left(\begin{array}{cccc}
                                            1 & 0 & 0 & 0 \\
                                            0 & \frac{1}{6} & 0 & 0 \\
                                            0 & 0 & \frac{1}{2} & 0 \\
                                            0 & 0 & 0 & 1
                                           \end{array}\right)_{J=3}$}}
                                           &\quad\quad
&$\langle\mathcal{E}_{7,9}[J]\rangle$
&\tiny{$\left(\begin{array}{cc}
                                   -\frac{\sqrt{\frac{3}{2}}}{2} & 0 \\
                                   0 & -\frac{\sqrt{\frac{3}{2}}}{2} \end{array}\right)_{J=1}$}
                           &\tiny{$\left(\begin{array}{ccc}
                                   -\frac{\sqrt{\frac{7}{2}}}{2}    & 0    & 0 \\
                                   0    & 0   & 0 \\
                               0     & 0     & -\frac{\sqrt{\frac{7}{2}}}{2}\end{array}\right)_{J=2}$}
                           &\scriptsize{$\left(\begin{array}{ccc}
                                   -1     & 0    & 0 \\
                                   0     & -\frac{\sqrt{\frac{3}{2}}}{2}     & 0 \\
                                   0     & 0   & -1\end{array}\right)_{J=3}$}\\
$\langle\mathcal{E}_{10}[J]\rangle$      &{$\left(\begin{array}{cc}
                                \frac{3}{2}    &0    \\
                                0      &\frac{3}{4}\end{array}\right)_{J=0}$}
                           &{$\left(\begin{array}{ccc}
                                             \frac{5}{4} & 0 & 0 \\
                                             0 & \frac{5}{4} & 0 \\
                                             0 & 0 & 0
                                            \end{array}\right)_{J=1}$}
                           &\scriptsize{$\left(\begin{array}{cccc}
                                             \frac{3}{4} & 0 & 0 & 0 \\
                                             0 & \frac{3}{2} & 0 & 0 \\
                                             0 & 0 & \frac{3}{4} & 0 \\
                                             0 & 0 & 0 & -1
                                            \end{array}\right)_{J=2}$}
                                            &\quad\quad
 &$\langle\mathcal{E}_{10}[J]\rangle$&\scriptsize{$\left(\begin{array}{ccc}
                                0    &0     &0\\
                                0    &\frac{5}{4}     &0\\
                                0    &0       &0\end{array}\right)_{J=3}$}
                            &\scriptsize{$\left(\begin{array}{ccc}
                                -1    &0      &0\\
                                0    &\frac{3}{4}     &0\\
                                0    &0      &-1\end{array}\right)_{J=4}$}\\
\bottomrule[1pt]\bottomrule[1pt]
\end{tabular}
\end{table*}

\renewcommand{\arraystretch}{0.8}
\begin{table*}[!htpb]
\caption{The matrix elements for the generalized tensor operators
$\mathcal{S}_i$ in the effective potentials. Here, the subscript $i$
of $\mathcal{S}_i$ corresponds to that in the effective potentials
in Eqs. (\ref{poten1}-\ref{poten11}).}\label{tensor}
\begin{tabular}{clll}\toprule[1pt]\toprule[1pt]
$\langle\mathcal{S}_i[J]\rangle$             &\multicolumn{3}{c}{Matrices}\\
\midrule[1pt]
$\langle\mathcal{S}_1[J]\rangle$        &{$\left(\begin{array}{cc}0  &\sqrt{2}\\
                             \sqrt{2}  &2\end{array}\right)_{J=0}$}
                            &{$\left(\begin{array}{cc}0           &-\sqrt{2}\\
                                     -\sqrt{2}   &1 \end{array}\right)_{J=1}$}
                            &{$\left(\begin{array}{ccc}
                                0    &\sqrt{\frac{2}{5}}    &-\sqrt{\frac{14}{5}}\\
                                \sqrt{\frac{2}{5}}    &0     &-\frac{2}{\sqrt{7}}\\
                                -\sqrt{\frac{14}{5}}    &-\frac{2}{\sqrt{7}}    &-\frac{3}{7}
                                \end{array}\right)_{J=2}$}\\
$\langle\mathcal{S}_{2,3}[J]\rangle$     &{$\left(\begin{array}{cc}0   &-\frac{2}{\sqrt{15}} \\
                                   -\frac{2}{\sqrt{15}}   &\sqrt{\frac{2}{15}}\end{array}\right)_{J=1}$}
                             &{$\left(\begin{array}{ccc}
                                  0    &\sqrt{\frac{3}{5}}      &0\\
                                  0    &0      &0\\
                                  0    &-\sqrt{\frac{6}{7}}    &0\end{array}\right)_{J=2}$}\\
$\langle\mathcal{S}_4[J]\rangle$        &{$\left(\begin{array}{cc}0  &\frac{1}{\sqrt{30}}\\
                                 \sqrt{\frac{7}{6}}  &\frac{1}{\sqrt{21}}\end{array}\right)_{J=0}$}
                             &{$\left(\begin{array}{cc}
                                0           &-\frac{1}{\sqrt{10}} \\
                                -\frac{1}{\sqrt{10}}     &\frac{1}{2\sqrt{5}}\end{array}\right)_{J=1}$}
                             &{$\left(\begin{array}{ccc}
                              0     &\sqrt{\frac{7}{30}}       &-\frac{1}{\sqrt{30}}\\
                                   \frac{1}{5\sqrt{6}}     &0       &-\frac{1}{\sqrt{105}}\\
                       -\frac{1}{\sqrt{30}}     &-\frac{1}{\sqrt{3}}     &-\frac{\sqrt{\frac{3}{7}}}{14}
                                       \end{array}\right)_{J=2}$}\\
$\langle\mathcal{S}_5[J]\rangle$
&{$\left(\begin{array}{cccc}
                    0 & \frac{3}{5 \sqrt{2}} & \sqrt{\frac{6}{5}} & \frac{\sqrt{\frac{21}{2}}}{5} \\
                    \frac{3}{5 \sqrt{2}} & -\frac{3}{10} & \sqrt{\frac{3}{5}} & -\frac{\sqrt{\frac{3}{7}}}{5} \\
                    \sqrt{\frac{6}{5}} & \sqrt{\frac{3}{5}} & \frac{1}{2} & \frac{2}{\sqrt{35}} \\
                    \frac{\sqrt{\frac{21}{2}}}{5} & -\frac{\sqrt{\frac{3}{7}}}{5} & \frac{2}{\sqrt{35}} & \frac{48}{35}
                   \end{array}\right)_{J=1}$}
                            &{$\left(\begin{array}{cccc}
                    0  & -\frac{3 \sqrt{2}}{5} & -\sqrt{\frac{7}{10}} & \frac{\sqrt{7}}{5} \\
                    -\frac{3 \sqrt{2}}{5}  & \frac{3}{10} & \frac{3}{\sqrt{35}} & -\frac{3 \sqrt{\frac{2}{7}}}{5} \\
                    -\sqrt{\frac{7}{10}}  & \frac{3}{\sqrt{35}} & -\frac{3}{14} & \frac{4 \sqrt{\frac{2}{5}}}{7} \\
                    \frac{\sqrt{7}}{5}  & -\frac{3 \sqrt{\frac{2}{7}}}{5} & \frac{4 \sqrt{\frac{2}{5}}}{7} & \frac{12}{35}
                   \end{array}\right)_{J=2}$}
                            &{$\left(\begin{array}{cccc}
                    0 & \frac{3}{5 \sqrt{2}} & -\frac{1}{\sqrt{5}} & -\frac{4 \sqrt{3}}{5} \\
                    \frac{3}{5 \sqrt{2}} & -\frac{3}{35} & -\frac{6 \sqrt{\frac{2}{5}}}{7} & -\frac{6 \sqrt{6}}{35} \\
                    -\frac{1}{\sqrt{5}} & -\frac{6 \sqrt{\frac{2}{5}}}{7} & -\frac{4}{7} & \frac{\sqrt{\frac{3}{5}}}{7} \\
                    -\frac{4 \sqrt{3}}{5} & -\frac{6 \sqrt{6}}{35} & \frac{\sqrt{\frac{3}{5}}}{7} & -\frac{22}{35}
                   \end{array}\right)_{J=3}$}\\
$\langle\mathcal{S}_6[J]\rangle$
&{$\left(\begin{array}{cccc}
                    0 & -\frac{23}{15 \sqrt{2}} & 2 \sqrt{\frac{2}{15}} & -\frac{\sqrt{\frac{7}{6}}}{5} \\
                    -\frac{23}{15 \sqrt{2}} & \frac{23}{30} & \frac{2}{\sqrt{15}} & \frac{1}{5 \sqrt{21}} \\
                    -2 \sqrt{\frac{2}{15}} & -\frac{2}{\sqrt{15}} & \frac{1}{2} & \frac{2}{\sqrt{35}} \\
                    -\frac{\sqrt{\frac{7}{6}}}{5} & \frac{1}{5 \sqrt{21}} & -\frac{2}{\sqrt{35}} & \frac{24}{35}
                   \end{array}\right)_{J=1}$}
                            &{$\left(\begin{array}{cccc}
                    0 & \frac{2 \sqrt{2}}{5} & -\sqrt{\frac{7}{10}} & \frac{\sqrt{7}}{5} \\
                    -\frac{2 \sqrt{2}}{5} & -\frac{23}{30} & \frac{2}{\sqrt{35}} & \frac{\sqrt{\frac{2}{7}}}{5} \\
                    -\sqrt{\frac{7}{10}} & -\frac{2}{\sqrt{35}} & -\frac{3}{14} & \frac{4 \sqrt{\frac{2}{5}}}{7} \\
                    -\frac{\sqrt{7}}{5} & \frac{\sqrt{\frac{2}{7}}}{5} & -\frac{4 \sqrt{\frac{2}{5}}}{7} & \frac{6}{35}
                   \end{array}\right)_{J=2}$}
                            &{$\left(\begin{array}{cccc}
                    0 & -\frac{1}{5 \sqrt{2}} & \frac{1}{\sqrt{5}} & -\frac{2 \sqrt{3}}{5} \\
                    -\frac{1}{5 \sqrt{2}} & \frac{23}{105} & -\frac{4 \sqrt{\frac{2}{5}}}{7} & \frac{2 \sqrt{6}}{35} \\
                    -\frac{1}{\sqrt{5}} & \frac{4 \sqrt{\frac{2}{5}}}{7} & -\frac{4}{7} & \frac{\sqrt{\frac{3}{5}}}{7} \\
                    -\frac{2 \sqrt{3}}{5} & \frac{2 \sqrt{6}}{35} & -\frac{\sqrt{\frac{3}{5}}}{7} & -\frac{11}{35}
                   \end{array}\right)_{J=3}$}\\
$\langle\mathcal{S}_{7,9}[J]\rangle$
&{$\left(\begin{array}{cc}
                                  0 & -\frac{2 \sqrt{3}}{5}  \\
                                  -\frac{2 \sqrt{3}}{5} & \frac{\sqrt{6}}{5} \end{array}\right)_{J=1}$}
                             &{$\left(\begin{array}{ccc}
                                  0         & 0   & -\frac{1}{\sqrt{5}} \\
                                \frac{3}{10}     & 0     & -\frac{3}{\sqrt{70}} \\
                                -\frac{1}{\sqrt{5}}    & 0    & -\frac{3}{7 \sqrt{14}}
                                \end{array}\right)_{J=2}$}
                             &{$\left(\begin{array}{ccc}
                                 0     & -\frac{3}{5 \sqrt{2}}        & -\frac{\sqrt{3}}{5} \\
                                 \frac{\sqrt{3}}{10}   & \frac{2 \sqrt{6}}{35}    & -\frac{6}{35}\\
                                -\frac{\sqrt{3}}{5} & \frac{6 \sqrt{6}}{35} &-\frac{11}{70}
                                \end{array}\right)_{J=3}$}\\
$\langle\mathcal{S}_{8}[J]\rangle$
&{$\left(\begin{array}{cccc}
                    0 & \frac{3}{5 \sqrt{2}} & -\sqrt{\frac{6}{5}} & \frac{\sqrt{\frac{21}{2}}}{5} \\
                    \frac{3}{5 \sqrt{2}} & -\frac{3}{10} & -\sqrt{\frac{3}{5}} & -\frac{\sqrt{\frac{3}{7}}}{5} \\
                    -\sqrt{\frac{6}{5}} & -\sqrt{\frac{3}{5}} & \frac{1}{2} & -\frac{2}{\sqrt{35}} \\
                    \frac{\sqrt{\frac{21}{2}}}{5} & -\frac{\sqrt{\frac{3}{7}}}{5} & -\frac{2}{\sqrt{35}} & \frac{48}{35}
                   \end{array}\right)_{J=1}$}
                             &{$\left(\begin{array}{cccc}
                    0  & \frac{3 \sqrt{2}}{5} & -\sqrt{\frac{7}{10}} & -\frac{\sqrt{7}}{5} \\
                    \frac{3 \sqrt{2}}{5}  & \frac{3}{10} & -\frac{3}{\sqrt{35}} & -\frac{3 \sqrt{\frac{2}{7}}}{5} \\
                    -\sqrt{\frac{7}{10}} & -\frac{3}{\sqrt{35}} & -\frac{3}{14} & -\frac{4 \sqrt{\frac{2}{5}}}{7} \\
                    -\frac{\sqrt{7}}{5} & -\frac{3 \sqrt{\frac{2}{7}}}{5} & -\frac{4 \sqrt{\frac{2}{5}}}{7} & \frac{12}{35}
                   \end{array}\right)_{J=2}$}
                             &{$\left(\begin{array}{cccc}
                    0 & \frac{3}{5 \sqrt{2}} & \frac{1}{\sqrt{5}} & -\frac{4 \sqrt{3}}{5} \\
                    \frac{3}{5 \sqrt{2}} & -\frac{3}{35} & \frac{6 \sqrt{\frac{2}{5}}}{7} & -\frac{6 \sqrt{6}}{35} \\
                    \frac{1}{\sqrt{5}} & \frac{6 \sqrt{\frac{2}{5}}}{7} & -\frac{4}{7} & -\frac{\sqrt{\frac{3}{5}}}{7} \\
                    -\frac{4 \sqrt{3}}{5} & -\frac{6 \sqrt{6}}{35} & -\frac{\sqrt{\frac{3}{5}}}{7} & -\frac{22}{35}
                   \end{array}\right)_{J=3}$}\\
$\langle\mathcal{S}_{10}[J]\rangle$
&{$\left(\begin{array}{cc}
                                 0    &\frac{3\sqrt{\frac{7}{10}}}{2}    \\
                                 \frac{3\sqrt{\frac{7}{10}}}{2}      &\frac{15}{14}
                                 \end{array}\right)_{J=0}$}
                             &{$\left(\begin{array}{ccc}
                                  0 & -\frac{13}{10 \sqrt{2}} & \frac{3 \sqrt{7}}{10} \\
                                  -\frac{13}{10 \sqrt{2}} & \frac{13}{20} & -\frac{3}{5 \sqrt{14}} \\
                                  \frac{3 \sqrt{7}}{10} & -\frac{3}{5 \sqrt{14}} & \frac{36}{35}
                                 \end{array}\right)_{J=1}$}
                             &{$\left(\begin{array}{cccc}
                                  0 & \frac{3 \sqrt{\frac{7}{2}}}{10} & -\frac{3 \sqrt{\frac{5}{14}}}{2} & \frac{9}{5 \sqrt{14}} \\
                                  \frac{3 \sqrt{\frac{7}{2}}}{10} & 0 & -\frac{3}{2 \sqrt{5}} & 0 \\
                                  -\frac{3 \sqrt{\frac{5}{14}}}{2} & -\frac{3}{2 \sqrt{5}} & -\frac{45}{196} & -\frac{18}{49 \sqrt{5}} \\
                                  \frac{9}{5 \sqrt{14}} & 0 & -\frac{18}{49 \sqrt{5}} & 0
                                 \end{array}\right)_{J=2}$}\\
$\langle\mathcal{S}_{10}[J]\rangle$&{$\left(\begin{array}{ccc}
                                0    &\frac{3\sqrt{3}}{10}     &-\frac{3\sqrt{3}}{5}\\
                                \frac{3\sqrt{3}}{10}    &\frac{13}{70}      &-\frac{18}{35}\\
                                -\frac{3\sqrt{3}}{5}    &-\frac{18}{35}      &-\frac{33}{70}
                               \end{array}\right)_{J=3}$}
                         &{$\left(\begin{array}{ccc}
                                0    &\frac{3}{\sqrt{70}}       &-\sqrt{\frac{11}{7}}\\
                               \frac{3}{\sqrt{70}}  &\frac{15}{49}  &-\frac{3\sqrt{\frac{55}{2}}}{49}\\
                                -\sqrt{\frac{11}{7}} &-\frac{3\sqrt{\frac{55}{2}}}{49}  &\frac{65}{98}
                             \end{array}\right)_{J=4}$}\\
$\langle\mathcal{S}_{11}[J]\rangle$
&{$\left(\begin{array}{cccc}
                    0 & -\frac{23}{15 \sqrt{2}} & -2 \sqrt{\frac{2}{15}} & -\frac{\sqrt{\frac{7}{6}}}{5} \\
                    -\frac{23}{15 \sqrt{2}} & \frac{23}{30} & -\frac{2}{\sqrt{15}} & \frac{1}{5 \sqrt{21}} \\
                    2 \sqrt{\frac{2}{15}} & \frac{2}{\sqrt{15}} & \frac{1}{2} & -\frac{2}{\sqrt{35}} \\
                    -\frac{\sqrt{\frac{7}{6}}}{5} & \frac{1}{5 \sqrt{21}} & \frac{2}{\sqrt{35}} & \frac{24}{35}
                   \end{array}\right)_{J=1}$}
                             &{$\left(\begin{array}{cccc}
                    0  & -\frac{2 \sqrt{2}}{5} & -\sqrt{\frac{7}{10}} & -\frac{\sqrt{7}}{5} \\
                    \frac{2 \sqrt{2}}{5}  & -\frac{23}{30} & -\frac{2}{\sqrt{35}} & \frac{\sqrt{\frac{2}{7}}}{5} \\
                    -\sqrt{\frac{7}{10}}  & \frac{2}{\sqrt{35}} & -\frac{3}{14} & -\frac{4 \sqrt{\frac{2}{5}}}{7} \\
                    \frac{\sqrt{7}}{5}  & \frac{\sqrt{\frac{2}{7}}}{5} & \frac{4 \sqrt{\frac{2}{5}}}{7} & \frac{6}{35}
                   \end{array}\right)_{J=2}$}
                             &{$\left(\begin{array}{cccc}
                    0 & -\frac{1}{5 \sqrt{2}} & -\frac{1}{\sqrt{5}} & -\frac{2 \sqrt{3}}{5} \\
                    -\frac{1}{5 \sqrt{2}} & \frac{23}{105} & \frac{4 \sqrt{\frac{2}{5}}}{7} & \frac{2 \sqrt{6}}{35} \\
                    \frac{1}{\sqrt{5}} & -\frac{4 \sqrt{\frac{2}{5}}}{7} & -\frac{4}{7} & -\frac{\sqrt{\frac{3}{5}}}{7} \\
                    -\frac{2 \sqrt{3}}{5} & \frac{2 \sqrt{6}}{35} & \frac{\sqrt{\frac{3}{5}}}{7} & -\frac{11}{35}
                   \end{array}\right)_{J=3}$}\\
\bottomrule[1pt]\bottomrule[1pt]
\end{tabular}
\end{table*}

%\vfill
%\newpage

\section*{ACKNOWLEDGMENTS}

This project is supported by the National Natural Science Foundation
of China under Grants No. 11222547, No. 11275115, No. 11175073, and
No. 11575008.

\end{document}